\documentclass[fleqn,usenatbib]{mnras}

\usepackage{newtxtext,newtxmath}
\usepackage{comment}
\usepackage{array}
\usepackage{float}
\usepackage{stfloats}
\usepackage{tabularray}
\usepackage{tabularx}
\usepackage{threeparttable}
\UseTblrLibrary{booktabs}
\usepackage{multirow}
\usepackage{mathtools}
\usepackage{adjustbox}

\usepackage[T1]{fontenc}
\usepackage{ae,aecompl}

\usepackage{graphicx}
\usepackage{amsmath}
\usepackage{amssymb}
\usepackage{enumitem}
\usepackage{xfrac}

\usepackage{xcolor}
\usepackage{hyperref}
\hypersetup{
    colorlinks=true,
    linkcolor=blue,
    citecolor=blue,
    urlcolor=blue
}

\setlist[itemize]{leftmargin=*}

\newcommand{\ud}{\mathrm{d}}

\newcommand{\vect}[1]{\boldsymbol{#1}}

\newcommand{\Msun}{M_\odot}

\newcommand{\arepo}{\texttt{AREPO}}

\newcommand{\ab}{a_{\mathrm{b}}}
\newcommand{\eb}{e_{\mathrm{b}}}
\newcommand{\Lb}{\vect{L}_{\mathrm{b}}}
\newcommand{\lhat}{\hat{\ell}_{\mathrm{b}}}
\newcommand{\adot}{\dot{a}_{\mathrm{b}}}
\newcommand{\edot}{\dot{e}_{\mathrm{b}}}
\newcommand{\Mdot}{\dot{M}}
\newcommand{\Rvisc}{R_{\mathrm{visc}}}

\newcommand{\Porb}{P_{\mathrm{orb}}}
\newcommand{\Enorm}{\mathcal{E}}
\newcommand{\alphavisc}{\alpha_{\mathrm{visc}}}

\title[Binary evolution in warped CBDs]{Alignment-driven binary evolution in warped circumbinary discs: 3D moving-mesh hydrodynamics of eccentric and inclined binaries}

\author[Souvaitzis et al.]{
Lazaros Souvaitzis$^{1,2}$\thanks{E-mail: lazaros.souvaitzis@mu.ie},
Magdalena Siwek$^{3}$
and Jorge Cuadra$^{4,5,6}$
\\
$^{1}$Max Planck Institute for Astrophysics, Karl-Schwarzschild-Str. 1, 85748 Garching, Germany\\
$^{2}$Centre for Astrophysics and Space Sciences Maynooth, Department of Physics, Maynooth University, Maynooth, Ireland\\
$^{3}$New York University, Center for Cosmology and Particle Physics, 726 Broadway, New York, NY 10003, USA\\
$^{4}$Departamento de Ciencias, Facultad de Artes Liberales, Universidad Adolfo Ib\'a\~nez, Av.\ Padre Hurtado 750, Vi\~na del Mar, Chile\\
$^{5}$Millennium Nucleus on Transversal Research and Technology to Explore Supermassive Black Holes (TITANS), Chile\\
$^{6}$Max-Planck-Institut f\"ur Extraterrestrische Physik, Gie{\ss}enbachstra{\ss}e, D-85748 Garching, Germany
}

\date{Accepted XXX. Received YYY; in original form ZZZ}
\pubyear{2026}

\begin{document}
\label{firstpage}
\pagerange{\pageref{firstpage}--\pageref{lastpage}}
\maketitle

\begin{abstract}
We present three-dimensional moving-mesh hydrodynamical simulations (\arepo{}) of equal-mass binaries in circumbinary discs, spanning binary eccentricity ($e = 0.0,\,0.3,\,0.6$) and inclination ($i = 0^\circ$--$60^\circ$), using a locally isothermal $\alpha$-disc ($H/r = \alphavisc = 0.1$) with fixed Keplerian orbits. Discs above the critical inclination $i_{\rm crit}(e)$, decreasing with $e$, drive their interiors near-polar, so eccentricity alone alters the outcome at fixed $i$. Discs reach polar alignment ten times faster than they align toward the binary plane, settling long before the binary's orbit responds. Alignment dictates how the disc acts on the binary. First, a disc tilting back toward the binary plane spends nearly all its torque reorienting the binary's orbit, whereas a disc driven polar redirects it into the orbit itself, which shrinks 10--20 times faster. Second, at $e = 0.6$ the turn to polar is a single transient reshaping the accretion flow, cavity and orbit, with accretion enhanced ($i = 45^\circ$) or suppressed ($i = 60^\circ$). Third, misalignment suppresses coplanar eccentricity pumping: the attractor at $e \approx 0.4$ survives only below $i \approx 15^\circ$, so misaligned binaries circularise and may enter the gravitational-wave regime less eccentric than coplanar models predict. Around a circular binary, the familiar $\approx 4.5$-orbit accretion modulation vanishes while the disc is strongly warped, returning as it realigns, a time-domain signature of misaligned massive black-hole binaries. Finally, observed polar discs around stellar binaries, such as HD~98800~BaBb's, are likely in their final state; intermediate misalignments reflect slower alignment toward the binary plane.
\end{abstract}

\begin{keywords}
accretion, accretion discs -- hydrodynamics -- binaries: general -- black hole physics -- gravitational waves -- methods: numerical
\end{keywords}

\section{Introduction}
\label{sec:intro}

Supermassive black-hole (SMBH) binaries are a generic outcome of hierarchical
galaxy formation: after a gas-rich merger the two nuclear black holes sink to the
centre of the remnant and form a bound pair \citep{Begelman1980}. Stellar
scattering and dynamical friction harden the binary to roughly parsec
separations, but in idealised, gas-poor spherical nuclei the hardening stalls
once the binary has ejected the stars on loss-cone orbits---the
``final-parsec problem'' \citep{Begelman1980,Milosavljevic2001}. More recent
work shows, however, that the binary keeps hardening in realistic merger remnants. In non-spherical
nuclei, collisionless loss-cone refilling sustains the hardening
\citep{Khan2011,Vasiliev2015,HolleyBockelmann2015}, and \citet{Gualandris2017}
find it efficient even for very mild triaxiality, concluding that stellar
dynamics alone can bring the binary to the gravitational-wave regime. Gas
\citep{Escala2005,Cuadra2009} and interactions with a third black hole
\citep{Bonetti2018} offer further routes. Whichever channel closes the final
parsec, gas shapes the late evolution of the binary.
A circumbinary disc (CBD), formed from the
cold gas funnelled to the nucleus during the merger \citep{Mayer2007,Cuadra2009},
exchanges mass, energy, and angular
momentum with the binary through resonant gravitational torques and accretion
\citep{Goldreich1980,Pringle1991,Artymowicz1994}; the gas can thereby drain (or supply)
orbital angular momentum and regulate how fast the binary
coalesces.   Because coalescing SMBH binaries are prime targets for the Laser
Interferometer Space Antenna \citep[LISA;][]{AmaroSeoane2017_LISA} and underlie
the nanohertz gravitational-wave background now reported by pulsar-timing arrays
\citep{Agazie2023}, their disc-driven evolution is a key ingredient of
multimessenger astrophysics. Identifying an electromagnetic counterpart is a central observational goal:
simulations post-processed with radiative transfer and ray-tracing predict
periodic accretion variability, distinctive spectral features and self-
lensing flares
\citep{Franchini2024,Cocchiararo2024,Krauth2024,Cocchiararo2025}, and the
observational case for sub-parsec SMBH binaries is reviewed by
\citet{DOrazioCharisi2023} and \citet{Krause2025}. The same binary--disc coupling shapes accreting binaries across the mass scale, most directly observed around young stellar and protostellar pairs \citep{Dutrey1994,Mathieu2000}, while high-resolution simulations of star-cluster formation show that circumbinary discs aid the hardening of massive binaries down to separations of 10~au \citep{Chon2026}.

The binary--disc coupling operates through three angular-momentum channels:
resonant gravitational torques exchanged with the cavity wall and spiral
arms, the direct accretion of gas and its momentum onto the components, and
--- for a misaligned disc --- an out-of-plane torque that reorients the
orbital plane rather than changing its size or shape.
Section~\ref{sec:theory_orbit} sets these out quantitatively; the first two
have been characterised extensively in two dimensions
\citep{Papaloizou1977, Artymowicz1996, ShiKrolik2012, ShiKrolik2015, Miranda2017, Munoz2019, Ohana2024, Most2024, Wang2025BMAD}, while the third
has no two-dimensional
counterpart and is one of the quantities this paper measures. Whether the binary spirals in (\emph{inspiral}) or out (\emph{outspiral}) depends on the angular momentum delivered per unit accreted mass, not only on the sign of the net exchange \citep{Goicovic2017}.

Early 2D simulations established the basic disc--binary phenomenology, from
the size of the cleared cavity \citep{Artymowicz1994} to its dependence on the
binary configuration \citep{DOrazio2013,Farris2014,DOrazio2016,Munoz2016}, but
simulations that excluded or poorly resolved the central cavity found the expected
binary inspiral \citep{MacFadyen2008,Cuadra2009}. Once the cavity gas was properly
captured, outspiral was first demonstrated by \citet{Miranda2017} and confirmed at
higher resolution with 2D \arepo{} by \citet{Munoz2019} (hereafter MML19), driven
by a net positive gravitational torque from the asymmetric, eccentric cavity wall;
independent 2D and 3D studies concurred
\citep{Moody2019,Duffell2020,Ragusa2020,Franchini2022,Franchini2023,Heath2020,Park2024}
and eleven codes were benchmarked on this setup \citep{Duffell2024}. Subsequent 2D surveys found a rich dependence on system properties: the
binary is
driven toward an eccentricity \emph{attractor} near $\eb\approx0.4$--$0.5$
\citep{Zrake2021,DOrazio2021,Siwek2023a}; mass ratio modulates both the net
expansion rate and preferential accretion onto the secondary
\citep{Duffell2020,Siwek2023b}; and the disc aspect ratio introduces a qualitative
switch, with outspiral reversing to inspiral below $h/r\lesssim0.04$
\citep{Tiede2020,Penzlin2022,Dittmann2024}, with further sensitivity to viscosity
\citep{Dittmann2022,Penzlin2022,Dittmann2024}. For a comprehensive review of these
results, see \citet{Lai2023}. Long-term semi-analytic integration confirms that disc
depletion by accretion typically outpaces orbital shrinkage, though eccentricity
excitation toward the attractor value remains a robust prediction
\citep{Valli2024}.

The transition to three dimensions introduces angular-momentum channels absent
from vertically-integrated models: \citet{Moody2019} found that vertical gas
motions weaken the cavity-edge gravitational torque, so the binary gains less
angular momentum than in 2D. \citet{Bourne2024} extended this 3D approach to
live-binary, self-gravitating discs, finding universal inspiral and, for misaligned
configurations at $i=45^\circ$, pronounced disc warping and breaking. Together, these results show that 3D effects change the torque budget qualitatively. The parameter space nevertheless remains largely unexplored: \citeauthor{Moody2019} considered only circular, coplanar
configurations, while \citeauthor{Bourne2024} operated in a self-gravitating regime
distinct from the locally-isothermal $\alpha$-disc baseline of the 2D surveys.

Whether the CBD is aligned with the binary orbit depends on how the gas
arrives. In gas-rich mergers the binary can form inside a massive nuclear disc
whose drag binds the black holes within a million years \citep{Mayer2007} and
brings them into corotation with the disc \citep{Dotti2009}, favouring
alignment. Gas delivered later, in chaotic accretion episodes or subsequent
mergers, instead reaches the binary carrying angular momentum poorly correlated
with the orbital plane \citep{Nixon2011}, and
simulations of turbulent clouds falling onto SMBH binaries form circumbinary and
circumsingle discs spanning a wide range of inclinations
\citep{Goicovic2016, Goicovic2017}. Recent simulations that follow the feeding
self-consistently from galactic scales down to the binary confirm that the
resulting circumbinary accretion is naturally warped, multiphase, and strongly
magnetised \citep{Wang2025}. Misalignment is therefore a common, if not
universal, outcome. It also changes how the gas acts on the binary. Sufficiently retrograde discs counter-align with the orbit, which
promotes rapid hardening \citep{Nixon2011}. Around eccentric binaries, strongly
inclined discs instead settle towards polar configurations, and tearing between
differentially precessing rings drops low-angular-momentum gas onto the binary on
dynamical rather than viscous time-scales; ejecting this gas by gravitational
slingshot offers a further route to shrinking the orbit \citep{Aly2015}. A systematic understanding of how a misaligned disc responds to the binary is
therefore essential.

The response of a misaligned disc to a central torque has been studied since
\citet{Bardeen1975} showed that Lense--Thirring precession, combined with
viscous dissipation, drives the inner regions of a disc into alignment with the
spin of a Kerr black hole while leaving a warp at the transition radius.
Three-dimensional simulations later showed that this warp need not remain
continuous: a sufficiently thin or inviscid disc tears into discrete rings instead \citep{Nealon2015, Raj2021}, and in accreting black-hole binaries the same
breaking limits how far the holes and disc can align \citep{Nealon2021}. A
binary produces this class of behaviour through its tidal quadrupole rather
than through frame dragging, and with a critical difference: the torque is
non-axisymmetric for an eccentric orbit, which opens the second alignment
outcome that organises this paper. When the disc plane is misaligned with the
binary orbit, the binary tidal torque
drives differential nodal precession across disc annuli. Whether the disc
responds as a rigid body or develops a warp depends on the ratio $\alpha/(H/r)$:
in the \emph{bending-wave} regime ($\alpha \lesssim H/r$) warp perturbations
propagate at $\approx c_{\rm s}/2$; in the \emph{diffusive} regime
($\alpha \gtrsim H/r$) they spread viscously \citep{Papaloizou1983, Ogilvie1999, Ogilvie2000, Lodato2010}. These laminar flows are themselves prone to a parametric instability of inertial waves \citep{Ogilvie2013, FairbairnOgilvie2023}, which competes with the  magnetorotational instability (MRI) in magnetised discs \citep{FairbairnStone2025}. When differential precession overwhelms the
disc's internal bending stress, the disc tears into independently-precessing rings;
in the diffusive regime the breaking condition depends on the disc aspect ratio,
viscosity, and inclination \citep{Nixon2012, Dogan2018}. For eccentric binaries, a
qualitatively new long-term equilibrium is accessible: discs misaligned beyond a
critical inclination $i_{\rm crit}(e_{\rm b})$ evolve toward a \emph{polar}
configuration in which the disc angular-momentum vector aligns with the binary
eccentricity vector rather than the orbital angular-momentum vector
\citep{Farago2010, MartinLubow2017, Cuello2019}. This polar alignment requires $e_{\rm b}>0$
and an initial misalignment $i_0 > i_{\rm crit}$; its dependence on binary
eccentricity, mass ratio, disc viscosity, size, and mass has been characterised
through simulations and analytic extensions \citep{Martin2018, Martin2019}.

Prior simulation studies of misaligned CBDs have been dominated by SPH
calculations \citep[e.g.][]{Nixon2013, Aly2015, MartinLubow2017, Martin2018}. SPH follows warped, misaligned discs naturally but
ties its resolution to the gas mass and its dissipation to an artificial
viscosity, while static grids resolve the low-density cavity at the cost of
advection errors where the flow is misaligned with the mesh; a moving Voronoi
mesh keeps the Lagrangian character of SPH while capturing the cavity with the
accuracy of a Godunov scheme \citep{Springel2010}. The first grid-based 3D study of an inclined disc
\citep{Fragner2010} established that thick, low-viscosity discs precess
rigidly while thinner or more viscous ones warp and break. With
\textsc{Athena++}, \citet{Rabago2023} found that high-$\alpha$ discs align
polar while low-$\alpha$ discs precess for long periods and develop cavity-
edge vortices, and \citet{Rabago2024} derived updated breaking criteria for
eccentric binaries. Spanning the full inclination range with unequal-mass eccentric
binaries, \citet{Yang2026Unequal} showed that the alignment outcome (coplanar vs.\ polar)
determines the preferential accretion fraction and that strong breaking episodes accelerate disc mass loss;
around eccentric binaries, the timing of the accretion pulses relative to
periastron also shifts with the misalignment \citep{Yang2026Pulsed}. For the binary orbital evolution, \citet{Chen2026} find
that polar-aligned CBDs universally drive binary inspiral, with decay rates
exceeding those of coplanar configurations. The live-binary simulations of
\citet{Bourne2024} found disc warping and breaking at $i=45^\circ$ in a
self-gravitating regime, with their inclined runs restricted to circular binaries. No 3D study
has combined the locally-isothermal $\alpha$-disc baseline of the 2D surveys with a
systematic grid in both binary eccentricity and disc inclination --- probing
simultaneously how eccentricity shifts $i_{\rm crit}$ and how proximity to the
polar-alignment boundary modifies the torque decomposition and disc morphology.

In this paper we present 3D \arepo{} simulations of an equal-mass binary
($q = m_2/m_1 = 1$) embedded in a locally-isothermal $\alpha$-disc
($H/r = 0.1$, $\alpha = 0.1$). A super-Lagrangian plus broken-power-law
refinement scheme maintains near-uniform resolution from the
binary's sub-orbital scale out to the transition radius $\Rvisc \approx 6\ab$,
resolving both the cavity-edge gas that sets the gravitational torque and the
warped outer disc. We hold the binary on a fixed Keplerian orbit, following the approach taken
in 2D by MML19, \citet{DOrazio2021} and \citet{Siwek2023a}, and in 3D by
\citet{Moody2019}, rather than the live-binary treatment of
\citet{Bourne2024};
this isolates the measurements of the gas torques at each fixed
$(e, i)$. The grid spans $e = 0.0,\,0.3,\,0.6$ and
$i = 0^\circ,\,15^\circ,\,30^\circ,\,45^\circ,\,60^\circ$ (15 production runs) and
addresses two interconnected questions. First: how does the disc's alignment
state --- tilt, warp and possible breaking --- respond to binary eccentricity
and inclination? Second: how does that alignment state govern the binary's
orbital response, including the out-of-plane torque channel that has no 2D
counterpart? We find that each disc settles into one of two alignment states,
tilting back toward the binary plane or turning polar, and that eccentricity
alone can select between them at fixed inclination. The discs warp smoothly
and none breaks. The alignment state in turn sets how the disc acts on the
binary: discs aligning toward the binary plane mainly reorient its orbit,
whereas polar discs act on the orbit's size and shape, and the turn to polar
is a single transient event --- a coupling between disc alignment, orbital
evolution and accretion variability captured by neither the 2D coevolution
nor the disc-breaking literature.

The paper is organised as follows. In Section~\ref{sec:theory} we provide the
governing equations for binary orbital evolution under gas torques and the
key disc physics relevant to warp propagation and disc alignment.
Section~\ref{sec:methods} describes the numerical setup, refinement scheme,
and diagnostic framework. Section~\ref{sec:disc_response} presents the disc response --- alignment, warping, and the $(e,i)$ landscape
--- and
Section~\ref{sec:orbital_evolution} the binary's orbital response and its coupling to the disc's alignment. We discuss the broader implications in
Section~\ref{sec:discussion} and summarise in Section~\ref{sec:conclusions}. The five
appendices provide the technical detail behind these results. Appendix~\ref{apndx:refinement} details the refinement scheme and the
resolution it delivers, on which every measurement relies.
Appendix~\ref{apndx:gallery} collects face-on density maps of all fifteen runs
and the accretion histories behind the mean rates of
Section~\ref{sec:disc_observability}. Appendix~\ref{apndx:transients} follows the two transient episodes in time: the partial decoupling of the $e = 0$, $i = 60^\circ$ disc and the turn to polar at $e = 0.6$, $i = 45^\circ$.
Appendix~\ref{apndx:detector} gives the test by which we establish that no disc
breaks.

\section{Theoretical Background}
\label{sec:theory}

\subsection{Binary orbital evolution under gas torques}
\label{sec:theory_orbit}

In code units ($G = M_{\mathrm{b}} = a_{\mathrm{b}} = 1$), the binary's specific
orbital energy and specific angular momentum are
\begin{equation}
\varepsilon_{\mathrm{b}} = -\frac{G M_{\mathrm{b}}}{2 a_{\mathrm{b}}}, \qquad
\ell_{\mathrm{b}}^2 = G M_{\mathrm{b}} a_{\mathrm{b}} (1 - e_{\mathrm{b}}^2).
\end{equation}
Differentiating with respect to time along the binary's orbit gives the
\citet{Munoz2019} forms (their Eqs.\ 34--35; identical content in
\citealt{DOrazio2021}),
\begin{equation}
\label{eq:adot}
\frac{\adot}{\ab} = \frac{\dot{M}_{\mathrm{b}}}{M_{\mathrm{b}}} - \frac{\dot{\varepsilon}_{\mathrm{b}}}{\varepsilon_{\mathrm{b}}},
\end{equation}
\begin{equation}
\label{eq:edot}
\edot = \frac{1 - e_{\mathrm{b}}^2}{2 e_{\mathrm{b}}}
        \left[ \frac{2\dot{M}_{\mathrm{b}}}{M_{\mathrm{b}}}
               - \frac{\dot{\varepsilon}_{\mathrm{b}}}{\varepsilon_{\mathrm{b}}}
               - \frac{2\dot{\ell}_{\mathrm{b}}}{\ell_{\mathrm{b}}} \right],
\end{equation}
where the three independent gas-torque inputs are the specific power
$\dot{\varepsilon}_{\mathrm{b}} = \vect{v}_{\mathrm{b}} \cdot \vect{f}_{\mathrm{ext}} - G\dot{M}_{\mathrm{b}}/r_{\mathrm{b}}$,
the specific angular-momentum rate
$\dot{\vect{\ell}}_{\mathrm{b}} = \vect{r}_{\mathrm{b}} \times \vect{f}_{\mathrm{ext}}$,
and the binary accretion rate
$\dot{M}_{\mathrm{b}} = \dot{M}_1 + \dot{M}_2$.
The external force per unit mass $\vect{f}_{\mathrm{ext}} = \vect{f}_{\mathrm{grav}} + \vect{f}_{\mathrm{acc}}$
includes both gravitational and accretion-recoil contributions.

In 3D, $\vect{\ell}_{\mathrm{b}}$ is a vector. The magnitude evolution that
enters Eqs.~\eqref{eq:adot}--\eqref{eq:edot} is the projection onto the
binary's angular-momentum direction,
\begin{equation}
\label{eq:ldot_proj}
\frac{\ud |\vect{\ell}_{\mathrm{b}}|}{\ud t} = \dot{\vect{\ell}}_{\mathrm{b}} \cdot \lhat,
\end{equation}
while torque components perpendicular to $\lhat$ change the \emph{direction} of
$\Lb$ (precession/tilt of the binary plane) without affecting $|\vect{\ell}_{\mathrm{b}}|$
and therefore do not enter the $a$--$e$ evolution.

\label{sec:theory_e0}%
For the nominally circular case ($e_{\mathrm{b}} = 0$), Eq.~\eqref{eq:edot} is
formally $0/0$. The well-defined limit is the bracket
\begin{equation}
\label{eq:bracket}
\mathcal{B} \equiv \frac{2\dot{M}_{\mathrm{b}}}{M_{\mathrm{b}}}
               - \frac{\dot{\varepsilon}_{\mathrm{b}}}{\varepsilon_{\mathrm{b}}}
               - \frac{2\dot{\ell}_{\mathrm{b}}}{\ell_{\mathrm{b}}},
\end{equation}
which equals $2 e_{\mathrm{b}} \edot$ as $e_{\mathrm{b}} \to 0$, making $\mathcal{B}$
the direct eccentricity-stability diagnostic for nominally circular runs.
A time-mean $\langle \mathcal{B} \rangle < 0$ implies that any residual eccentricity
is damped back to zero --- the circular attractor is stable --- while
$\langle \mathcal{B} \rangle > 0$ signals an unstable fixed point
\citep[cf.][their Fig.~1]{DOrazio2021}. This $e_{\mathrm{b}} = 0$ fixed point
coexists with a separate \emph{eccentric} attractor near $e_{\mathrm{b}} \approx
0.4$--$0.5$, toward which binaries with larger initial eccentricities are driven
\citep{Zrake2021, DOrazio2021, Siwek2023a}; the two are separated by an unstable
point at $e_{\mathrm{b}} \approx 0.07$ in 2D \citep[][their Fig.~1]{DOrazio2021}.
Our $e_{\mathrm{b}} = 0$ runs probe the stability of the lower, circular fixed point
in 3D.

\subsection{Evolution of warped discs}
\label{sec:theory_warp}

The response of a misaligned disc to a binary's tidal torque depends on the
relative magnitudes of viscosity and aspect ratio. For $\alpha < H/r$ the disc
communicates warps as bending waves at speed $c_{\mathrm{bw}} \approx c_{\mathrm{s}}/2$
\citep{Papaloizou1983}, while for $\alpha > H/r$ warp propagation is diffusive.
Our disc parameters ($\alpha = H/r = 0.1$) sit on the boundary between these
two regimes \citep{Ogilvie1999}. The $c_{\mathrm{bw}} \approx c_{\mathrm{s}}/2$
estimate refers to the laminar internal flows that carry the warp; these flows are
themselves subject to a hydrodynamic parametric instability of inertial waves whose
non-linear saturation can reduce the effective warp-propagation efficiency \citep{Ogilvie2013, FairbairnOgilvie2023}, and which competes with the magnetorotational
instability in magnetised discs \citep{FairbairnStone2025}. We will be working with the
laminar estimate in the rest of this paper.

The local amplitude of a warp is measured by the dimensionless warp parameter
\begin{equation}
\label{eq:psi}
\psi(r,t) \;=\; r\left|\frac{\partial \hat{\vect{l}}_{\mathrm d}}{\partial r}\right|
\;=\; \left|\frac{\partial \hat{\vect{l}}_{\mathrm d}}{\partial \ln r}\right| ,
\end{equation}
the reorientation of the disc's angular-momentum direction per unit
logarithmic radius \citep{Ogilvie1999}. It is the natural non-linearity
parameter of warp theory: the linear (small-warp) treatments of warp
propagation assume $\psi \ll 1$, warps with $\psi$ approaching unity drive
internal flows fast enough to modify the transport itself
\citep{Dogan2018}, and local shearing-box calculations identify
$\psi \gtrsim 1$ as the threshold above which a warped disc becomes unstable
and can break \citep{Held2026}. Section~\ref{sec:warp_no_break} measures $\psi$
for every run in our grid and Section~\ref{sec:disc_regime} places our results
against this threshold.

\subsubsection{The warp radius}
\label{sec:theory_warp_radius}

Both criteria that follow are statements about a single characteristic
radius. The external torque that drives differential precession and the
internal torque that communicates the warp scale differently with radius, so
their ratio passes through unity at a \emph{warp radius} $r_{\rm w}$,
\begin{equation}
\label{eq:warp_radius}
\mathcal{A}(r) \;\equiv\; \frac{G_{\rm c}(r)}{G_{\rm p}(r)} ,
\qquad \mathcal{A}(r_{\rm w}) = 1 ,
\end{equation}
where $G_{\rm p}$ is the precession torque per unit area exerted on the disc
by the binary's quadrupole and $G_{\rm c}$ the torque per unit area by which
the disc communicates the warp. The precession torque falls off with radius
faster than the communication torque, so $\mathcal{A} < 1$ inside
$r_{\rm w}$ and $\mathcal{A} > 1$ outside: inside it the disc reorients onto
the orientation the binary sets, outside it retains its original
inclination, and the transition between the two is the warp
\citep{Lodato2006, Tremaine2014}. Which channel carries the warp decides whether such a radius exists at all.
In the diffusive regime ($\alpha > H/r$) the communicating agent is the
vertical viscosity $\nu_2$ --- the transport coefficient acting on the
vertical shear of a warp, distinct from the $\nu_1$ that drives accretion
and related to it by $\nu_2 \simeq \nu_1/2\alpha^2$ for small warps \citep{Papaloizou1983, Lodato2010} --- and $r_{\rm w}$ follows from balancing it against
the local precession rate. In the wave-like regime ($\alpha < H/r$)
bending waves carry the tilt across the disc faster than viscosity can damp
it, and the disc precesses as a unit rather than developing a viscous
transition: there is no warp radius \citep{Kao2026}. In that regime the steady solutions carry a radial oscillation of the tilt, whose inner amplitude can exceed the outer tilt several-fold \citep{Drewes2021}. Such an oscillation does not define a transition radius, so the absence of $r_{\rm w}$ there is a property of the wave-like regime itself. Our discs sit on that boundary, so we treat $r_{\rm w}$ as a quantity to be
measured rather than predicted (Section~\ref{sec:warp_no_break}).

\subsubsection{Tearing criterion for circular binaries}
\label{sec:theory_warp_nixon}

In the diffusive regime, \citet{Nixon2013} showed that a circumbinary disc
tears where the quadrupole precession torque exceeds the viscous torque that
communicates the warp. For a circular binary this yields a tearing radius
\begin{equation}
\label{eq:r_break_nixon}
r_{\mathrm{break}} \lesssim \left[ \frac{\mu \, |\sin 2i|}{4\,\alpha}
                    \, \frac{r}{H} \right]^{1/2} \ab,
\end{equation}
where $\mu = m_2/M_{\mathrm{b}}$ \citep[cf.\ the warp-instability
generalisation of][]{Dogan2018}. The criterion depends on inclination --- vanishing toward $i = 0^\circ$ and
$90^\circ$ --- but not on binary eccentricity. Equation~\eqref{eq:r_break_nixon} is Eq.~\eqref{eq:warp_radius} with $G_{\rm c}$
taken to be the viscous torque alone. \citet{Nixon2013} note that this estimate neglects
pressure effects, which come to dominate in thicker discs where wave
propagation matters --- the regime our $\alpha = H/r$ discs border.
For our parameters ($\mu = 0.5$, $\alpha = H/r = 0.1$)
Eq.~\eqref{eq:r_break_nixon} gives $r_{\mathrm{break}} \approx 2.5$--$3.5\,\ab$
across the inclinations of our grid, largest at $i = 45^\circ$ where
$|\sin 2i|$ peaks. A break \emph{would} therefore sit well inside the disc body,
and in every run inside the measured warp radius, $r_{\mathrm{break}} < r_{\rm w}$,
if the disc were able to break. None of our discs does
(Section~\ref{sec:warp_no_break}).

\subsubsection{Nodal precession and the wave-like breaking criterion}
\label{sec:theory_warp_rabago}

\citet{Rabago2024} derived an updated breaking criterion using global disc
timescales that remain valid for extended discs (unlike the thin-ring limit
implicit in Nixon--Do\u{g}an). A misaligned ring at radius $r$ is driven by the
binary's quadrupole to precess about whichever axis governs its alignment, at the local rate $k\,(\ab/r)^{7/2}\,\Omega_{\mathrm{b}}$, so an internally communicating
disc precesses as a unit at the angular-momentum-weighted mean of those rates,
\begin{equation}
\label{eq:omega_d}
\Omega_{\mathrm{d}} = k \left\langle \left(\frac{\ab}{r}\right)^{7/2}
\right\rangle \Omega_{\mathrm{b}},
\qquad
\left\langle \left(\frac{\ab}{r}\right)^{7/2} \right\rangle =
\frac{\displaystyle\int_{r_{\mathrm{in}}}^{r_{\mathrm{out}}}
      \Sigma\, r^3 \Omega\, (\ab/r)^{7/2}\,\ud r}
     {\displaystyle\int_{r_{\mathrm{in}}}^{r_{\mathrm{out}}}
      \Sigma\, r^3 \Omega\,\ud r},
\end{equation}
where $\Omega = \sqrt{G M_{\mathrm{b}}/r^3}$ is the Keplerian angular frequency
and the weight $\Sigma r^3 \Omega$ is the local angular-momentum content
\citep[][their eqs.~17 and 19]{Yang2026Unequal}. The dimensionless precession
coefficient $k$ is set by the binary:
\begin{equation}
\label{eq:k}
k = \begin{cases}
-\dfrac{3}{4}\sqrt{1+3e_{\mathrm{b}}^2 - 4e_{\mathrm{b}}^4}\,
\dfrac{M_1 M_2}{M_{\mathrm{tot}}^2}
& \text{(coplanar disc),} \\[8pt]
\dfrac{3}{4}\sqrt{5}\,e_{\mathrm{b}}\sqrt{1+4e_{\mathrm{b}}^2}\,
\dfrac{M_1 M_2}{M_{\mathrm{tot}}^2}
& \text{(polar disc).}
\end{cases}
\end{equation}
For a power-law disc ($\Sigma \propto r^{-p}$) the average in
Eq.~\eqref{eq:omega_d} evaluates in closed form and returns exactly the global
precession time $t_{\mathrm{p}} = 1/|\Omega_{\mathrm{d}}|$ tabulated by
\citet{Rabago2024}; the two formulations are the same quantity. We work with
the integral form because it can be evaluated on the \emph{measured} surface-density
profile of each run instead of an assumed power law, which is how we place our
runs on the breaking landscape in Section~\ref{sec:disc_regime}.
A disc breaks where this precession time falls below the time
$t_{\mathrm{c}} \simeq [4/(s+2)]\,(r_{\mathrm{out}}^{s/2+1} -
r_{\mathrm{in}}^{s/2+1})$ required for bending waves to communicate across it
($c_{\mathrm{s}}^2 \propto r^{-s}$); the breaking radius is the outer radius at which $t_{\mathrm{p}} =
t_{\mathrm{c}}$, beyond which the disc cannot communicate the precession to
its outer edge before tearing. This is Eq.~\eqref{eq:warp_radius} once more,
with the warp carried by bending waves and the balance struck globally
rather than locally. Breaking in this regime has also been found in simulations: \citet{Drewes2021} obtain tearing in three-dimensional simulations of a disc
warped by a central spinning mass rather than by a binary, with the total
viscosity measured to satisfy $\alpha < H/r$ at the moment the disc breaks.
At $q = 1$, the factor $M_1 M_2 / M_{\mathrm{tot}}^2 = 1/4$ is maximised, making
equal-mass binaries the most efficient at driving disc breaking. For polar alignment, the precession rate exceeds the coplanar rate for $e_{\mathrm{b}} > 0.41$:
our $e = 0.6$ runs lie in this regime, so on precession-rate grounds alone they
would be the most break-prone members of the grid. One limitation of the two-case coefficient $k$
should be noted: it describes precession well away from the critical
inclination, whereas close to the dividing surface the precession period
diverges and neither rate applies.

\subsubsection{Viscous suppression of breaking at $\alpha = H/r$}
\label{sec:theory_warp_viscous}

Whether a disc can actually break also depends on whether viscous alignment acts
faster than precession. \citet{Rabago2024} compare the viscous alignment time
$t_{\mathrm{align}} = \alpha^{-1}(H/r)^2\,\Omega_{\mathrm{b}}/\Omega_{\mathrm{d}}^2$
\citep[see also][]{Yang2026Unequal} with the precession time
$t_{\mathrm{p}} = 1/|\Omega_{\mathrm{d}}|$ through the dimensionless ratio
\begin{equation}
\label{eq:tau_visc}
\tau \equiv \frac{t_{\mathrm{align}}}{t_{\mathrm{p}}} =
\frac{1}{\alpha}\left(\frac{H}{r}\right)^{2}
\frac{\Omega_{\mathrm{b}}}{|\Omega_{\mathrm{d}}|}.
\end{equation}
When
$\tau < 1$, the disc aligns with the local precession vector faster than it can
complete a precession cycle, suppressing large misalignments between adjacent rings
and \emph{preventing a break}. \citeauthor{Rabago2024}'s grid-based simulations
break the disc only for $\alpha \lesssim H/r$; their $\alpha = 0.1 = H/r$ run
\emph{warps but does not break}. Our discs share that
viscosity and aspect ratio, and Section~\ref{sec:disc_regime} compares them with
the criterion.

\subsubsection{Coplanar versus polar alignment: the critical inclination}

\begin{figure}
\includegraphics[width=\columnwidth]{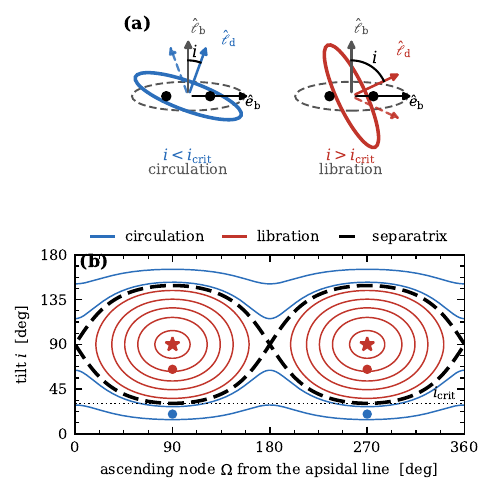}
\caption{Nodal circulation about $\lhat$ versus libration about $\hat{\vect{e}}_{\rm b}$. \textit{(a)} A disc tilted by $i$ from the binary plane, with its ascending node on the apsidal meridian ($\Omega = 90^\circ$) as in our initial conditions: below $i_{\rm crit}$ the node circulates about $\lhat$ (blue); above it the node librates about $\hat{\vect{e}}_{\rm b}$ and the disc is driven polar (red). Dashed arrows show a second position of $\hat{\vect{l}}_{\rm d}$ on each precession cone, mirrored through its axis. \textit{(b)} Level sets of the secular energy $\Enorm$ (Eq.~\eqref{eq:ehat}) at $e_{\rm b} = 0.6$: the black dashed separatrix crosses $\Omega = 90^\circ$ at $i_{\rm crit}$ (Eq.~\eqref{eq:icrit}; dotted line); stars mark the polar fixed points and dots the two discs of panel (a).}
\label{fig:circlib}
\end{figure}

\label{sec:theory_warp_icrit}

\citet{Nixon2013} predict that breaking is
most efficient for $45^\circ < i < 135^\circ$ and strongest at $i = 45^\circ$ and
$135^\circ$ --- a result derived for circular binaries and Lense--Thirring precession.
The two alignment behaviours are dynamically distinct, and
Fig.~\ref{fig:circlib} shows both. Around a \emph{circular} binary a
misaligned ring undergoes nodal \emph{circulation}, its angular momentum
$\hat{\vect{l}}_{\mathrm d}$ precessing about $\lhat$ at nearly constant
tilt \citep{Nixon2012, Facchini2013, Foucart2013}. Around an
\emph{eccentric} binary a second regime opens above a critical tilt: the
precession axis shifts to the binary eccentricity vector
$\hat{\vect{e}}_{\mathrm b}$ and the node \emph{librates} rather than
circulating \citep{Verrier2009, Farago2010, Doolin2011, Aly2015}. With
dissipation the first damps toward coplanar prograde alignment and the
second toward polar \citep{MartinLubow2017, Martin2018, Yang2026Unequal}. The two
are separated by an energy surface of the quadrupole secular problem
\citep[][their \S2.4 and fig.~1]{Farago2010}; writing that energy in the
form used by \citet{Aly2015}, we characterise each disc by the normalised
energy
\begin{equation}
\label{eq:ehat}
\Enorm \;=\; \frac{E}{15\,e_{\mathrm{b}}^2 + 3\,(1 - e_{\mathrm{b}}^2)} ,
\end{equation}
where
\begin{equation}
\label{eq:esecular}
E \;=\; 15\,e_{\mathrm{b}}^2
    (\hat{\vect{l}}_{\mathrm d}\cdot\hat{\vect{e}}_{\mathrm b})^2
  \;-\; 3\,(1 - e_{\mathrm{b}}^2)
    (\hat{\vect{l}}_{\mathrm d}\cdot\lhat)^2 ,
\end{equation}
rings with $E < 0$ circulate about $\lhat$ and rings with $E > 0$ librate about
$\hat{\vect{e}}_{\mathrm b}$, with $E = 0$ the dividing surface. The denominator --- the value for a perfectly polar disc plus the magnitude
of the coplanar value --- maps
$\Enorm$ onto $[-1, +1]$ with the dividing surface at exactly zero for every
$e_{\mathrm{b}}$, so that the coordinate is comparable across eccentricities.
Two limits recur below: the polar region centre
($\hat{\vect{l}}_{\mathrm d} = \hat{\vect{e}}_{\mathrm b}$) sits at $\Enorm
= +0.738$ for $e_{\mathrm{b}} = 0.6$, while the coplanar state
($\hat{\vect{l}}_{\mathrm d} = \lhat$) sits at $-1$, $-0.67$ and $-0.26$ for
$e_{\mathrm{b}} = 0$, $0.3$ and $0.6$: the polar state lies energetically closer to a coplanar disc at higher
eccentricity, which is Eq.~\eqref{eq:icrit} expressed
as an energy. At $e_{\mathrm{b}} = 0$ there is no apsidal line and the first
term is absent.
\citet{Rabago2024} show that this critical inclination
$i_{\mathrm{crit}}(e_{\mathrm{b}})$ also organises disc breaking: near
$i_{\mathrm{crit}}$ the precession rate $|\omega_{\mathrm{p}}|$ passes
through zero, making breaking \emph{unlikely} in that neighbourhood, while breaking is favoured at intermediate inclinations within each of the
two basins of attraction, coplanar and polar. For our eccentric runs the critical inclination follows
\begin{equation}
i_{\mathrm{crit}} \;=\; \arctan\!\left(
\sqrt{\frac{1 - e_{\mathrm{b}}^2}{5\,e_{\mathrm{b}}^2}}\,\right)
\label{eq:icrit}
\end{equation}
\citep{Farago2010, Aly2015, MartinLubow2017}, giving $i_{\mathrm{crit}} \approx 54.9^\circ$ for
$e = 0.3$ and $i_{\mathrm{crit}} \approx 30.8^\circ$ for $e = 0.6$.
Equation~\eqref{eq:icrit} is the test-particle (massless-disc) criterion ---
the dividing surface of the outer restricted problem
($k^2 = 1$, equivalently $h = e_{\mathrm{b}}^2$, in the notation of
\citealt{Farago2010}, their \S2.4 and eq.~2.30) --- recast as an inclination
for an initial node on the apsidal meridian ($\phi_0 = \pm 90^\circ$, where
the polar window is widest), the geometry our initial conditions realise. The massless limit is the self-consistent one for our
prescribed binary orbit; generalisations that include the disc-to-binary
angular-momentum ratio \citep[e.g.][]{Martin2019, Yang2026Unequal} apply to
live binaries and massive discs. Our grid ($i \leq 60^\circ$) samples the
coplanar-prograde and polar attractors; retrograde coplanar alignment (approached from $i \gtrsim 140^\circ$) lies outside its range
\citep[for coplanar retrograde discs in 2D, see][]{ONeill2026}. The $i = 45^\circ$ pair therefore falls on opposite sides of the separatrix,
the $e = 0.6$ disc additionally benefiting from the $e_{\mathrm{b}} > 0.41$
precession boost of Eq.~\eqref{eq:k}, and the measured tilt evolution
follows (Section~\ref{sec:branches}). Eccentricity alone therefore reverses the alignment outcome, and with it the
binary's torque radial distribution (Section~\ref{sec:disc_response}).

\section{Methods}
\label{sec:methods}

\subsection{Numerical method}
\label{sec:methods_code}

We use the moving-mesh code \arepo{} \citep{Springel2010, Pakmor2016} in its
Navier--Stokes (explicit viscous) configuration \citep{Munoz2013}, which solves the
compressible viscous flow equations on an unstructured Voronoi tessellation advected
with the gas. The disc and binary follow the locally-isothermal, viscous \arepo{}
set-up of \citet{Munoz2019} and \citet{Siwek2023a, Siwek2023b}, extended here to
three dimensions. We work in code units $G = M_{\mathrm{b}} = \ab = 1$, so that the
binary orbital period is $\Porb = 2\pi$.

\subsection{Disc model}
\label{sec:methods_disc}

The equation of state is locally isothermal, with the pressure $P = \rho\,c_{\mathrm{s}}^2$
set each step from a position-dependent sound speed
\begin{equation}
\label{eq:soundspeed}
c_{\mathrm{s}}(\vect{r}) = h\,\sqrt{|\Phi_{\mathrm{b}}(\vect{r})|},
\end{equation}
where $\Phi_{\mathrm{b}}$ is the (softened) binary potential and $h \equiv H/R = 0.1$
is the fixed aspect ratio. Far from the binary $|\Phi_{\mathrm{b}}| \to G M_{\mathrm{b}}/R$,
so $c_{\mathrm{s}} \propto R^{-1/2}$ and $H/R = 0.1$ is constant throughout the disc.
The binary components are sink particles with softening length
$b_{\mathrm{soft}} = 0.04\,\ab$ and accretion radius $r_{\mathrm{acc}} = 0.05\,\ab$.

The initial disc follows the locally-isothermal, viscous steady-state profile of
\citet{Munoz2020}, with surface density
\begin{equation}
\label{eq:sigma}
\Sigma(R) = \Sigma_0 \left(\frac{R}{\ab}\right)^{-1/2}
            \left[1 - 0.7\left(\frac{\ab}{R}\right)^{1/2}\right]
            f_{\mathrm{cav}}(R)\, f_{\mathrm{out}}(R),
\end{equation}
where the $R^{-1/2}$ scaling is the viscous steady-state solution
($\Sigma \propto \dot{M}/\nu$) and $f_{\mathrm{cav}}$, $f_{\mathrm{out}}$ are tapering
functions setting an inner cavity at $R_{\mathrm{in}} = 2\,\ab$ and an outer edge at
$R_{\mathrm{out}} = 50\,\ab$. We extend this to three dimensions by distributing the
surface density in a Gaussian vertical layer of scale height $H(R) = h R$,
\begin{equation}
\label{eq:rho3d}
\rho(R, z) = \rho_0(R)\,\exp\!\left[-\frac{z^2}{2 H(R)^2}\right],
\qquad \rho_0(R) = \frac{\Sigma(R)}{H(R)},
\end{equation}
so that the disc keeps the constant aspect ratio $h = 0.1$ at every radius. The
normalisation $\Sigma_0 = 10^{-4}\,M_{\mathrm{b}}\,\ab^{-2}$ sets a low-mass disc
whose self-gravity is neglected, so that the disc mass enters only as an overall
scaling of the gas torques and accretion rate. The gas is initialised on
near-Keplerian orbits with corrections for the binary quadrupole moment and the
radial pressure gradient, together with a viscous radial-drift velocity
\citep[following][]{Munoz2019}; the central cavity is then carved self-consistently
by the binary tides, with no non-axisymmetric perturbation imposed.

For the viscosity we adopt the $\alpha$-disc prescription of
\citet{Siwek2023a, Siwek2023b}. For a single accretor the kinematic viscosity is
$\nu = \alpha\,h\,c_{\mathrm{s}}\,R$ with $\alpha = 0.1$; for the binary the
contributions of the two components are blended so that each circumsingle disc is
itself an $\alpha$-disc of the same aspect ratio,
\begin{equation}
\label{eq:viscosity}
\nu = \alpha\,h\,c_{\mathrm{s}}\,
      \frac{R_1 R_2}{\left(R_1^{\,n} + R_2^{\,n}\right)^{1/n}},
\qquad n = 5,
\end{equation}
where $R_1$, $R_2$ are the distances to the two sinks. This recovers
$\nu \to \alpha\,h\,c_{\mathrm{s}}\,R_k$ within the Roche lobe of component $k$ and
the single-disc value $\nu \to \alpha\,h\,c_{\mathrm{s}}\,R$ in the circumbinary disc
at $R \gg \ab$.

\subsection{Binary configuration}
\label{sec:methods_binary}

The binary is placed on a fixed Keplerian orbit (the gas torques are diagnosed
but not applied to the orbit) with equal masses $m_1 = m_2 = 0.5\,M_{\mathrm{b}}$,
semi-major axis $\ab = 1$, and eccentricity $e_{\mathrm{b}} \in \{0.0, 0.3, 0.6\}$.
Equal masses are an idealisation: in coplanar 2D discs, equal-mass binaries at
$e_{\mathrm{b}} \approx 0.2$--$0.3$ accrete away from $q = 1$ rather than
holding it \citep{DeLaurentiis2026}, so our $e_{\mathrm{b}} = 0.3$ runs fix a
mass ratio that gas accretion would not preserve.
The orbital plane is inclined at $i \in \{0^\circ, 15^\circ, 30^\circ, 45^\circ, 60^\circ\}$
with respect to the initial disc mid-plane, implemented as a rotation about the
$y$-axis. The disc mid-plane is initially aligned with the $x$--$y$ plane.
Because the binary orbit is fixed, any apparent precession of $\Lb$ in our
diagnostics is a response in the disc, not binary precession.

The resulting fifteen production runs are referred to as
\[
\mathrm{e}\left\{\begin{array}{c} 00 \\ 03 \\ 06 \end{array}\right\}
\mathrm{i}\left\{\begin{array}{c} 0 \\ 15 \\ 30 \\ 45 \\ 60 \end{array}\right\},
\]
where the first bracket gives $e_{\mathrm{b}}$ in tenths and the second $i$ in
degrees; for example, e06i45 denotes $e_{\mathrm{b}} = 0.6$, $i = 45^\circ$.
All runs use $q = 1$, $H/r = \alpha = 0.1$ and a fixed orbit, and every run was
integrated to $t = 1100\,\Porb$, the common endpoint of the analysis windows
used throughout this paper. The conclusions are insensitive to that endpoint:
recomputing the cross-run statistics at successively earlier endpoints changed
no verdict and no sign, although individual amplitudes moved.

\subsection{Refinement}
\label{sec:methods_refinement}

Resolving the cavity and cavity-wall gas --- where the binary--disc torque
coupling is strongest --- while keeping the cell count tractable across a disc
extending to $50\,\ab$ requires more than standard quasi-Lagrangian
refinement, which holds the cell \emph{mass} fixed and therefore severely
under-resolves the low-density cavity. We impose instead a geometric
\emph{super-Lagrangian} target cell volume set by proximity to the sinks and
to the binary barycentre, following \citet{Curtis2015, Curtis2016} and \citet{Bourne2024}. The prescription is seam-free, and delivers
$N_z \equiv H/r_{\mathrm{cell}} \gtrsim 4$ cells per scale height across the
cavity wall and inner disc, never falling below $N_z \approx 1.3$ in the disc
body, with ${\sim}5$ cells across the accretion radius. The full
prescription, its parameters and the resulting resolution profile are given in
Appendix~\ref{apndx:refinement}.

\subsection{Diagnostics}
\label{sec:methods_diagnostics}

Our orbital-evolution diagnostics are built from the gas forces on the
binary. At every synchronisation point we accumulate, separately for each
component, the gravitational force from all gas and the force carried by
accreted material, together with the corresponding rate of work and torque,
splitting the gravitational part into gas inside and outside the binary
separation. Summing over the components gives the specific power
$\dot{\varepsilon}_{\mathrm{b}}$, the specific angular-momentum rate
$\dot{\vect{\ell}}_{\mathrm{b}}$ and the accretion rate
$\dot{M}_{\mathrm{b}}$ of Eqs.~\eqref{eq:adot}--\eqref{eq:edot}, and hence
$\adot/\ab$, $\edot$ and, for the circular runs, the bracket $\mathcal{B}$
(Eq.~\eqref{eq:bracket}). Because the torque is a vector we track its
projection onto $\lhat$, which drives $\adot$ and $\edot$, separately from
its perpendicular components, which reorient the orbital plane
(Section~\ref{sec:theory_orbit}). Note that the perpendicular torque
\emph{vector} itself lies in the binary's orbital plane --- a torque
perpendicular to $\vect{\ell}_{\mathrm{b}}$ tips $\vect{\ell}_{\mathrm{b}}$
--- so ``out-of-plane'' labels its \emph{effect}, not its direction. Writing
\begin{equation}
\dot{\ell}_{\mathrm{b},\parallel} = \dot{\vect{\ell}}_{\mathrm{b}}\cdot\lhat,
\qquad
\dot{\vect{\ell}}_{\mathrm{b},\perp} = \dot{\vect{\ell}}_{\mathrm{b}}
 - \dot{\ell}_{\mathrm{b},\parallel}\,\lhat,
\label{eq:tsplit}
\end{equation}
we quote the out-of-plane fraction
\begin{equation}
\frac{|T_\perp|}{|T|} \;=\;
\frac{\bigl|\langle \dot{\vect{\ell}}_{\mathrm{b},\perp} \rangle_w\bigr|}
{\sqrt{\langle \dot{\ell}_{\mathrm{b},\parallel} \rangle_w^2
 + \bigl|\langle \dot{\vect{\ell}}_{\mathrm{b},\perp} \rangle_w\bigr|^2}},
\label{eq:tperp_ratio}
\end{equation}
where $\langle\cdot\rangle_w$ is the secular window mean, taken as a
\emph{vector} mean for the perpendicular part so that orbital-timescale
oscillations cancel. We also verified the bookkeeping by reconstructing the
gravitational force on each sink from snapshot gas distributions with
\arepo{}'s own softened kernel, recovering the time-integrated on-the-fly
forces to $|\Delta f|/|f| \sim 10^{-5}$.

A second on-the-fly diagnostic resolves that torque spatially, supplying the
radial decomposition of Section~\ref{sec:orbital_origin} without
post-processing. About twenty times per orbit we bin the gravitational torque
and power into $100$ logarithmic bins in cylindrical radius over
$0.1$--$50\,\ab$, integrated over height, and the torque alone into $16$
linear bins in $z/H(R)$ over $\pm 4$ scale heights, integrated over radius;
we call this the $(R,z)$ accumulator. Because these are instantaneous samples
at that cadence rather than running integrals, a window mean over them is a
true orbit average, where a roughly per-orbit snapshot cadence would sample
the binary stroboscopically. Summed over all radial bins it reproduces the
independently computed binary gravitational torque to $0.09$ per cent. A
complementary snapshot-based reconstruction resolves the same torque by gas
height above the orbital plane, and yields the 3D-to-2D torque ratio of
Section~\ref{sec:orbital_origin}.

To characterise the disc structure we decompose the gas angular momentum into
concentric shells in each cell's Keplerian semi-major axis about the binary
barycentre (the binary treated as a point mass $M_{\mathrm b}$; unbound gas
excluded) and follow the orientation of each shell's
angular-momentum unit vector $\hat{\vect{l}}_{\mathrm d}(r,t)$. For gas on
near-circular orbits this coordinate coincides with radius, and we write it
as $r$ throughout, reserving $R$ for cylindrical radius; the two differ
appreciably only inside the cavity, where the gas moves on eccentric streams. The angle
between a shell and the binary plane gives the tilt profile $\delta(r,t)$,
the differential reorientation between adjacent shells gives the warp
amplitude $\psi(r,t)$ (Eq.~\eqref{eq:psi}), and the azimuth of the shell
angular momentum --- referenced to the binary eccentricity vector for the
eccentric runs --- tracks the disc's nodal precession.

\section{Disc Response: Alignment and Warping}
\label{sec:disc_response}

Across the fifteen runs of the $e \times i$ grid the disc response organises
into a dichotomy set by the critical inclination $i_{\rm crit}(e)$ of
Section~\ref{sec:theory_warp_icrit}: discs launched below it damp gently
toward the binary plane, while those launched above it --- the $e=0.6$ runs
at $i=45^\circ$ and $60^\circ$, against $i_{\rm crit} = 30.8^\circ$ ---
drive their interiors to a near-polar orientation. Every disc in the grid warps smoothly; \emph{none} tears, in
agreement with the viscous suppression expected at $\alpha = H/r$
\citep{Rabago2024}. The single marginal case is the steepest warp of all ($e=0$, $i=60^\circ$),
which transiently reaches partial decoupling near $t \approx 464\,P_{\rm b}$
before relaxing (Section~\ref{sec:warp_no_break}).

\begin{table}
\centering
\caption{Disc and binary timescales of the twelve inclined runs, in units of $10^3\,\Porb$. $t_{1/2}$: time for the inner disc ($r = 2$--$4\,\ab$) to halve its mutual inclination with the binary or, for the two polar discs, its offset from the polar orientation (reached only transiently at $e = 0$, $i = 60^\circ$). $P_{\rm prec,b}$: precession period of the binary's orbital plane about the disc's angular momentum. $t_i = i/|\dot{i}|$, $t_e = \eb/|\edot|$ and $t_a = \ab/|\adot|$, with $i$ measured from the whole disc. The semi-major axis shrinks in every run; $e$ grows only at $e = 0.3$, $i = 15^\circ$, and $i$ only in the two polar runs. Binary timescales are measured over $t = 500$--$1100\,\Porb$ for the simulated disc and scale inversely with its mass (or accretion rate), whereas $t_{1/2}$ does not.}
\label{tab:timescales}
\footnotesize
\setlength{\tabcolsep}{4pt}
\begin{tabular}{ccccccc}
\toprule
 & & disc & \multicolumn{4}{c}{binary} \\
\cmidrule(lr){3-3}\cmidrule(lr){4-7}
$e$ & $i$ & $t_{1/2}$ & $P_{\rm prec,b}$ & $t_i$ & $t_e$ & $t_a$ \\
\midrule
0.0 & $15^\circ$ & $>1.1$ & 17 & 5.0 & -- & 550 \\
0.3 & $15^\circ$ & $>1.1$ & 11 & 4.2 & 7.5 & 31 \\
0.6 & $15^\circ$ & $>1.1$ & 7.6 & 7.5 & 17 & 140 \\
\midrule
0.0 & $30^\circ$ & $>1.1$ & 16 & 4.0 & -- & 120 \\
0.3 & $30^\circ$ & $>1.1$ & 10 & 3.3 & 32 & 26 \\
0.6 & $30^\circ$ & $>1.1$ & 6.0 & 4.3 & 3.4 & 33 \\
\midrule
0.0 & $45^\circ$ & $>1.1$ & 13 & 2.7 & -- & 44 \\
0.3 & $45^\circ$ & $>1.1$ & 8.7 & 2.4 & 1.9 & 19 \\
0.6 & $45^\circ$ & 0.31 & 50 & 11 & 0.82 & 2.5 \\
\midrule
0.0 & $60^\circ$ & 0.42 & 24 & 4.7 & -- & 51 \\
0.3 & $60^\circ$ & $>1.1$ & 9.6 & 2.7 & 0.52 & 12 \\
0.6 & $60^\circ$ & 0.12 & 77 & 18 & 0.93 & 3.2 \\
\bottomrule
\end{tabular}
\end{table}

\subsection{Alignment states}
\label{sec:branches}

\begin{figure*}
\includegraphics[width=\textwidth]{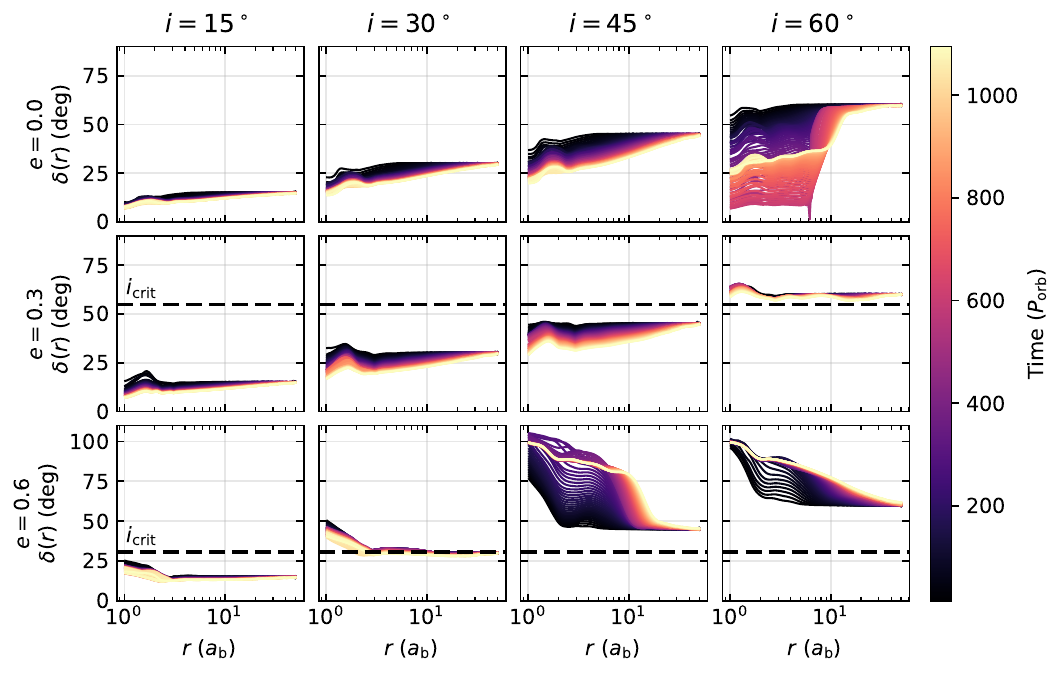}
\caption{Disc tilt profiles $\delta(r)$ (Eq.~\eqref{eq:delta_def}) relative to the binary orbital plane, window-averaged and coloured by time (rows: eccentricity; columns: $i = 15^\circ$--$60^\circ$). The dashed line marks $i_{\rm crit}(e)$ (Eq.~\eqref{eq:icrit}): $54.9^\circ$ at $e = 0.3$ and $30.8^\circ$ at $e = 0.6$ (at $e = 0$ it lies at $90^\circ$ and is omitted). Discs that align toward the binary plane keep nearly rigid, gently damping profiles. The polar discs ($e=0.6$, $i=45^\circ$ and $60^\circ$) drive their interiors to $\delta \approx 88^\circ$ while the outer disc holds its initial tilt; their innermost rings ($r \lesssim 2\,\ab$) librate past $90^\circ$. The $e=0$, $i=60^\circ$ disc concentrates its alignment front at $r \approx 6$--$9\,\ab$ (Section~\ref{sec:warp_no_break}).}
\label{fig:delta_grid}
\end{figure*}

\begin{figure*}
\centering
\includegraphics[height=0.42\textheight]{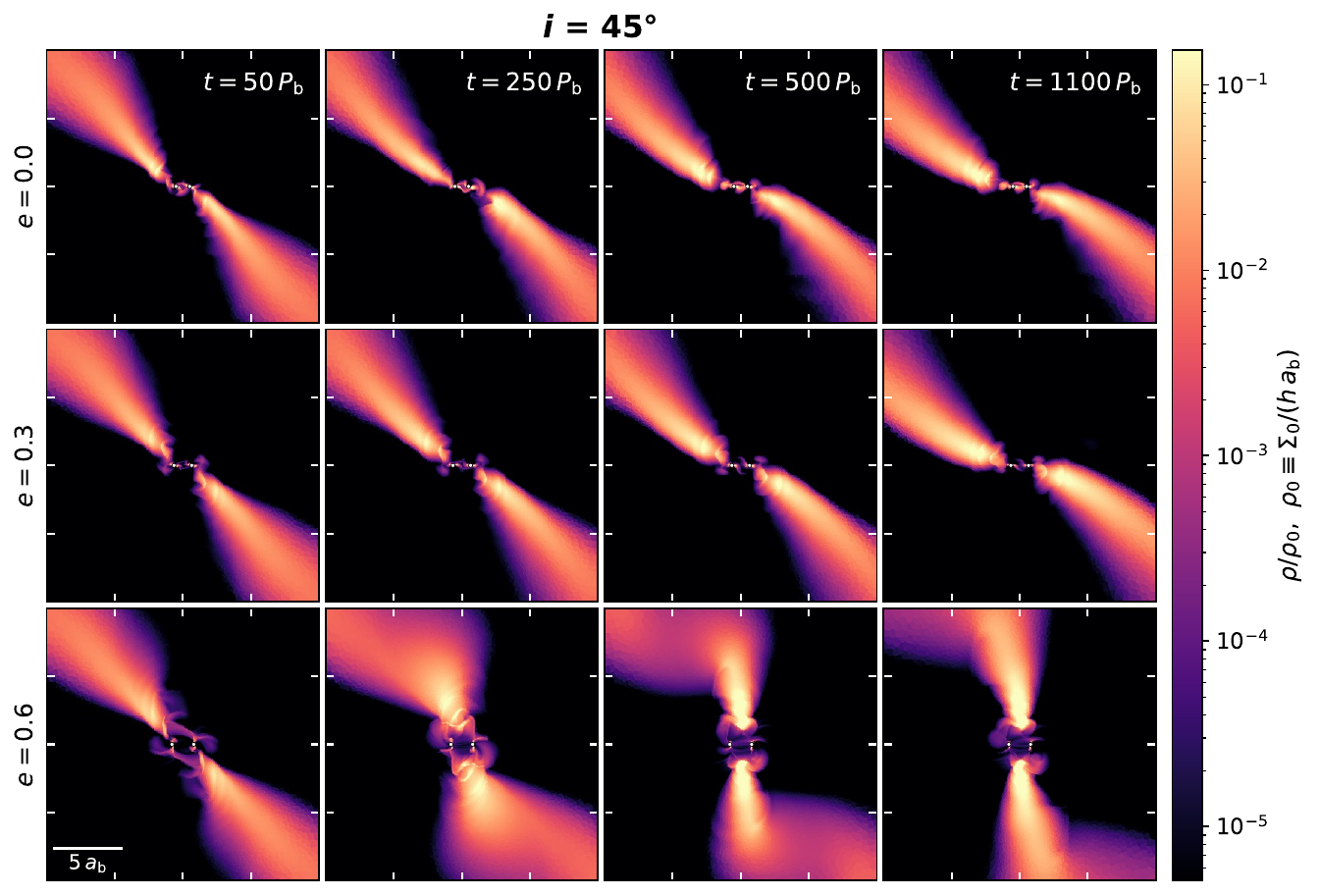}\\[3pt]
\includegraphics[height=0.42\textheight]{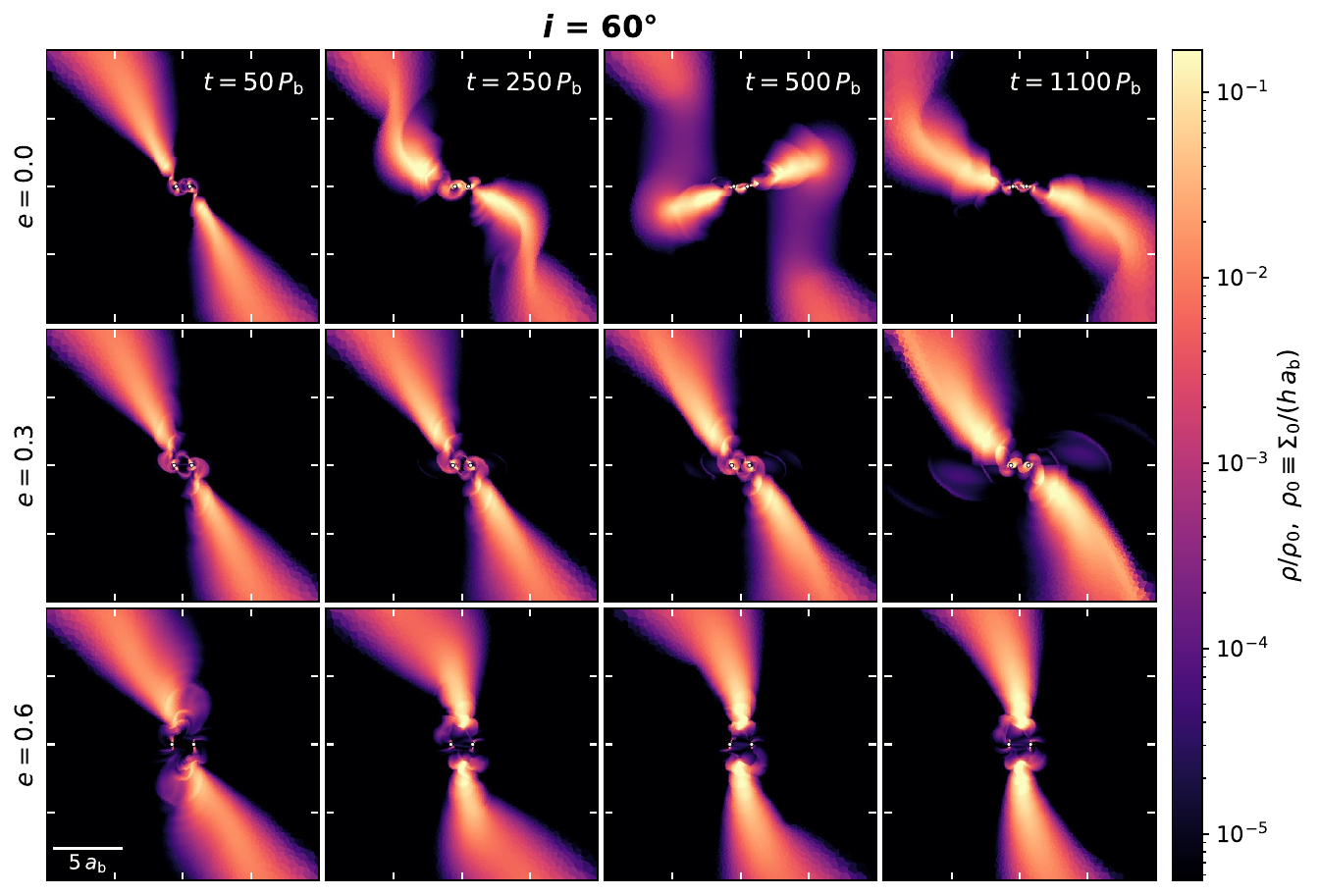}
\caption{Edge-on ($x$--$z$) density cuts of the $i=45^\circ$ (top block) and $i=60^\circ$ (bottom block) discs, with rows $e = 0,\,0.3,\,0.6$ and columns $t = 50,\,250,\,500$ and $1100\,\Porb$, in the binary frame: the binary plane is horizontal, so a polar disc appears vertical. The aligning and near-critical discs remain coherent, inclined bands, except $e=0$, $i=60^\circ$, whose flattened interior separates from its inclined exterior (the partial decoupling of Section~\ref{sec:warp_no_break}), sharpest at $500\,\Porb$ and gone by $1100\,\Porb$. The polar discs ($e=0.6$) turn their interiors vertical, caught mid-turn at $250\,\Porb$ for $i=45^\circ$. White dots mark the sinks; each block has its own colour scale.}
\label{fig:mosaic}
\end{figure*}

\begin{figure}
\includegraphics[width=\columnwidth]{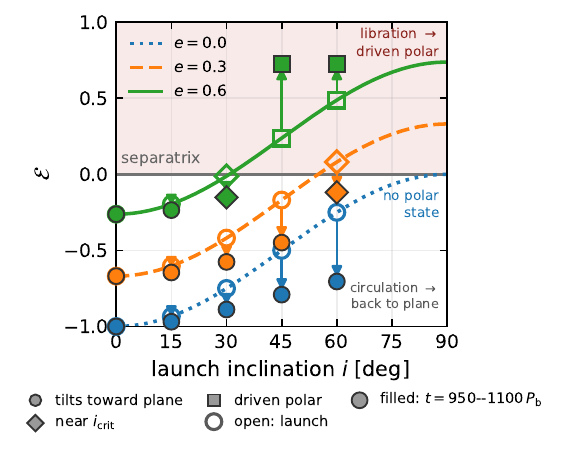}
\caption{Where the fifteen runs start and end relative to the separatrix: secular energy $\Enorm$ (Eq.~\eqref{eq:ehat}) against launch inclination for each eccentricity; open symbols mark the launch value, filled symbols the value over $t = 950$--$1100\,\Porb$. Above the separatrix $\Enorm = 0$ (shaded)
discs librate about $\hat{\vect{e}}_{\rm b}$ and are driven polar; below it
they circulate about $\lhat$ and tilt back toward the binary plane. The
curves cross it at $i_{\rm crit} = 55^\circ$ ($e = 0.3$) and $31^\circ$
($e = 0.6$); at $e = 0$ only at $i = 90^\circ$. The near-critical runs
(diamonds) both end just below the separatrix, one of them after launching
just above it.}
\label{fig:separatrix}
\end{figure}

Figure~\ref{fig:delta_grid} shows the radial tilt profiles
\begin{equation}
\delta(r,t) \;=\; \arccos\!\bigl[\hat{\vect{l}}_{\mathrm d}(r,t)\cdot\lhat\bigr],
\label{eq:delta_def}
\end{equation}
window-averaged and coloured by time (the shell vectors
$\hat{\vect{l}}_{\mathrm d}$ are defined in
Section~\ref{sec:methods_diagnostics}; all cross-run statistics use a common
analysis window ending at $t = 1100\,P_{\rm b}$). Discs comfortably below the critical inclination --- every $e=0$
run, $e=0.3$ at $i \leq 45^\circ$, and $e=0.6$ at $i=15^\circ$ --- precess about
$\lhat$ as nearly rigid, mildly warped bodies: their tilt profiles settle within
$\sim$100 orbits and then damp slowly toward alignment, from the inside out. The
steepest such case is $e=0$, $i=60^\circ$, whose interior descends from $60^\circ$ to a minimum of $\approx 9^\circ$
near $t \approx 594\,P_{\rm b}$ before tilting back while the disc
beyond $r \approx 8\,a_{\rm b}$ still holds the initial tilt --- an alignment
front that concentrates the full initial misalignment into a few scale
lengths. The $e=0.6$, $i=45^\circ$ disc is qualitatively different: its interior ($r
\lesssim 6\,a_{\rm b}$) climbs from $45^\circ$ to $\delta \approx 88^\circ$
while the exterior stays near its initial inclination. The interior settles
just short of the exactly polar orientation ($\delta = 90^\circ$, with
$\hat{\vect{l}}_{\mathrm d}$ parallel to $\hat{\vect{e}}_{\mathrm b}$) because it
remains coupled, through the warp, to an outer disc that still holds its
initial tilt. The $i=60^\circ$ counterpart repeats that reorientation faster, by $t \approx 115\,P_{\rm b}$. The two runs launched within a few degrees of $i_{\rm crit}$ bracket it, and
behave differently from one another. At $e=0.6$, $i=30^\circ$ ($i - i_{\rm
crit} = -0.8^\circ$) the disc stays quiescent at its initial tilt. At
$e=0.3$, $i=60^\circ$ ($+5.1^\circ$) it divides: the outer disc stays on the
libration side of the separatrix while the innermost annuli cross to
circulation (Fig.~\ref{fig:separatrix}), placing the effective hydrodynamic polar-alignment boundary
slightly \emph{above} the analytic $i_{\rm crit}$. Figure~\ref{fig:mosaic} shows the
corresponding density evolution for the $i=45^\circ$ and $i=60^\circ$ discs
in the binary frame, where the alignment structure is directly visible: the
$e \leq 0.3$ discs remain inclined, coherent planes at every epoch, while the
$e=0.6$ interiors rotate to a vertical --- near-polar --- orientation as
their cavities contract, between $t \approx 150$ and $250\,P_{\rm b}$
(reaching near-polar by $t \approx 290\,P_{\rm b}$) at $i=45^\circ$ and
already by $t \approx 115\,P_{\rm b}$ at $i=60^\circ$. Because
$i_{\rm crit}(0.3) = 54.9^\circ$ but $i_{\rm crit}(0.6) = 30.8^\circ$, the two $i=45^\circ$ discs fall on opposite sides of it: \emph{eccentricity
alone changes the alignment outcome at fixed inclination}
\citep{Farago2010, MartinLubow2017}.

Figure~\ref{fig:separatrix} places every run's launch and late-time state
against the analytic prediction of Eq.~\eqref{eq:icrit}, and the measured
outcomes confirm it throughout:
the two discs launched well above $i_{\rm crit}$ reorient their interiors to
$\delta \approx 88^\circ$, the two launched within a few degrees of it ---
one just below, one just above --- remain within $2^\circ$ of their initial
tilt, and the remaining eight damp
toward the binary plane. The signed distance from the boundary, $i - i_{\rm
crit}$, orders the observed transition speeds: the disc furthest onto polar alignment reorients fastest,
while the run sitting at the boundary ($e = 0.6$, $i = 30^\circ$) remains
quiescent. Near the separatrix the secular precession period grows without
bound, so this disc barely precesses within the simulated time; with little
differential precession to drive a warp and dissipate it, its tilt hardly
evolves.

Table~\ref{tab:timescales} puts numbers on these transition speeds; its binary columns are discussed in Section~\ref{sec:orbital_maps}. The inner discs ($r = 2$--$4\,\ab$) that
tilt toward the binary plane reduce their mutual inclination with the binary by 14--45 per cent by $t = 1100\,\Porb$, so none halves it within the run except, briefly, the transiently decoupling $e = 0$, $i = 60^\circ$ disc; where the decline is close to exponential, extrapolating it gives a half-time of $(1.8$--$2.4) \times 10^3\,\Porb$. Polar alignment is an order of magnitude faster: the two discs
driven polar halve their angular offset from the polar orientation in $124$ and $312\,\Porb$. In both channels the measured time lies within a factor of
about three of the viscous alignment time of
Section~\ref{sec:theory_warp_viscous} evaluated at the inner disc
($t_{\rm align} \approx 100$ and $120\,\Porb$ for the polar pair,
$(1.2$--$3.3) \times 10^3\,\Porb$ otherwise), so the analytic estimate
carries over to our 3D discs to that accuracy. The disc as a whole barely moves, covering only 2--16 per cent of the angle to its end state. Nor does the inner disc
keep precessing at its local rate ($t_{\rm p} \approx 12$--$73\,\Porb$): its
node shifts by $25$--$100^\circ$ early on and then stalls. Apart from the
$e = 0$, $i = 60^\circ$ disc, whose inner node regressed through more than a
full cycle during its transient partial decoupling, no run shows late-time
precession faster than $t_{\rm p} \sim 10^3\,\Porb$.

\subsection{Disc Dynamics Across Eccentricity and Inclination}
\label{sec:phasemap}

\begin{figure*}
\includegraphics[width=\textwidth]{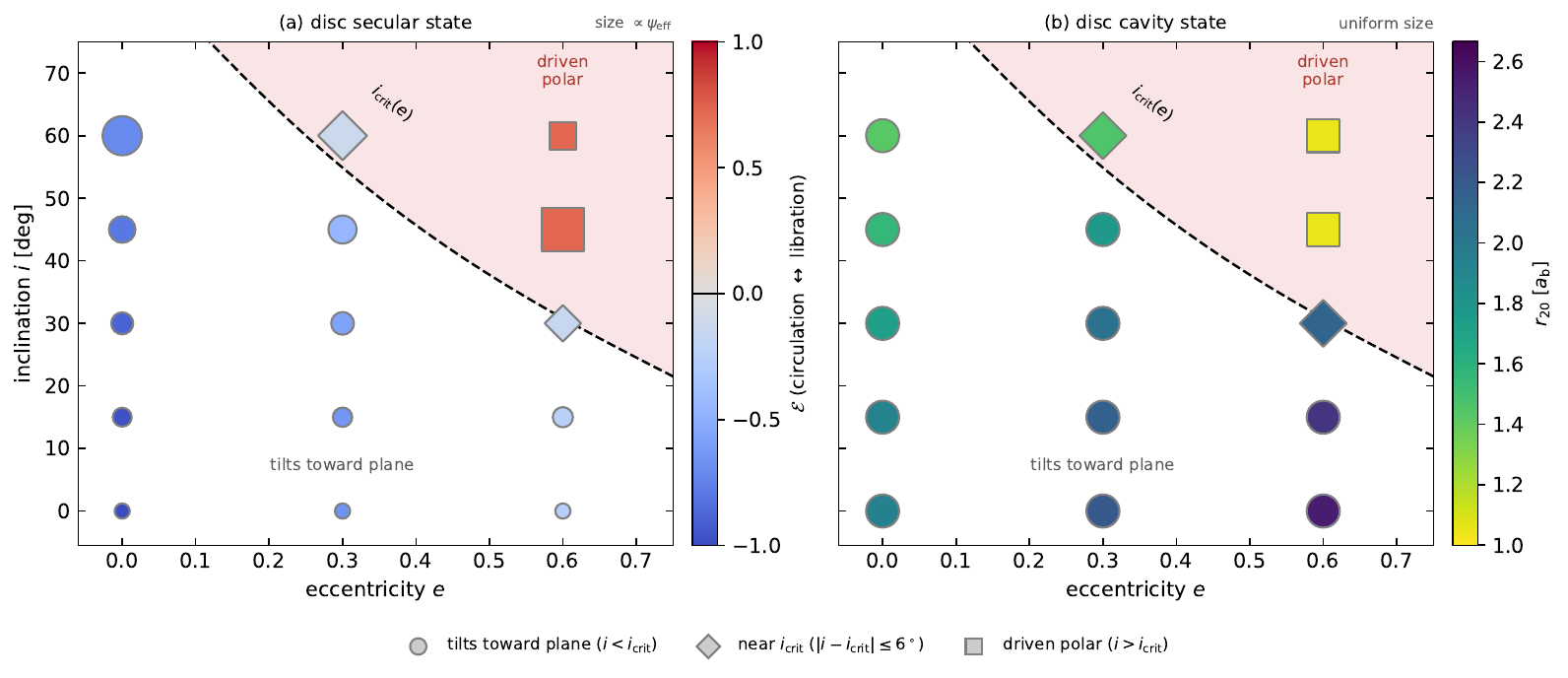}
\caption{The disc-side $(e,i)$ landscape. \textit{(a)} Normalised secular energy $\Enorm$ of the inner disc (Eq.~\eqref{eq:ehat}), from the $|L|$-weighted mean direction of the $r = 2$--$4\,\ab$ shells over $t = 950$--$1100\,\Porb$, on a diverging scale centred on the separatrix $\Enorm = 0$ (blue: circulation about $\lhat$; red: libration about $\hat{\vect{e}}_{\rm b}$); symbol size scales with the warp amplitude $\psi_{\rm eff}$ (Eq.~\eqref{eq:psieff}). \textit{(b)} Late-time cavity size $\langle r_{20}\rangle$ ($t = 1040$--$1100\,\Porb$). Symbol shape gives the predicted alignment class (circle: $i < i_{\rm crit}$, tilts toward the binary plane; diamond: within $6^\circ$ of $i_{\rm crit}$; square: $i > i_{\rm crit}$, driven polar); the dashed curve is $i_{\rm crit}(e)$ \citep{Farago2010}, with the polar side shaded.}
\label{fig:phase_map}
\end{figure*}

Figure~\ref{fig:phase_map} assembles the disc-side outcomes in the $(e,i)$
plane: the orientation each inner disc settles into (panel a) and the cavity
it maintains (panel b). Panel (a) colours each run by $\Enorm$
(Eq.~\eqref{eq:ehat}), which measures how far a disc sits from the dividing
surface $\Enorm = 0$: negative values belong to discs that circulate about
$\lhat$ and tilt back toward the binary orbital plane, positive values to
discs that librate about $\hat{\vect{e}}_{\rm b}$ and are driven polar. It measures both which of the two a disc settles into and how far. The two discs that turn polar reach $\Enorm = +0.73$ against a maximum
of $+0.738$ for a perfectly polar orientation at $e = 0.6$: they are as far
into the polar state as this configuration permits. The
two runs launched closest to the critical inclination remain closest to
the dividing surface ($-0.12$ at $e = 0.3$, $i = 60^\circ$ and $-0.15$ at
$e = 0.6$, $i = 30^\circ$), which is why neither reorients
appreciably. The remainder lie in the circulating region, moving closer to the dividing
surface as either eccentricity or inclination increases.

The $i = 0^\circ$ runs isolate the mechanism. With no misalignment at all,
$\Enorm$ still rises from $-1.00$ to $-0.67$ to $-0.26$ as eccentricity
goes from $0$ to $0.3$ to $0.6$: a disc around a more eccentric binary
begins closer to the dividing surface even when perfectly aligned, so
eccentricity has already carried the disc most of the way toward that
surface before any tilt is applied. The
cavity panel tracks the same structure through a different observable: the
cavity grows with eccentricity at low inclination, shrinks with inclination
within each column, and reaches its smallest radii in the two polar discs,
whose reoriented interiors no longer meet the binary in its own plane. This
reproduces the effect \citet{Franchini2019} proposed as an observational
diagnostic: resonant torques on the inner disc weaken in the polar
configuration, so a polar disc around an eccentric binary carves a smaller
cavity than a coplanar one, and the observed inner radius can therefore
constrain the disc's inclination --- the argument by which they confirmed
polar alignment in HD~98800~BaBb. We measure the cavity by $r_{20}$, the smallest radius at which the azimuthally
averaged surface density rises through one fifth of its peak value, taken on
the binary-plane profile built from cells within $|z| \leq 1\,\ab$; the
search is restricted to $r \geq 1\,\ab$ so that the circumsingle discs
cannot set the peak, making $r_{20}$ a circumbinary cavity wall by
construction. Angle brackets denote a plain average over the snapshots of
the quoted window, and the polar pair reaches $\langle r_{20}\rangle \approx
1.05\,\ab$.

\subsection{Smooth disc warping}
\label{sec:warp_no_break}

The warp in each inclined disc sits at the warp radius $r_{\rm w}$ of
Section~\ref{sec:theory_warp_radius}, measured as the radius of largest
misalignment between the disc interior and exterior (Appendix~\ref{apndx:detector} gives the method and defines the warp amplitude $\psi_{\rm eff}$). It lies in the disc body, at
$r_{\rm w} = 6.2$--$12.4\,\ab$ (Table~\ref{tab:disc_response}). Where the
warp can be followed in time it migrates outward as the disc communicates its
tilt (Appendix~\ref{apndx:transients}), so $r_{\rm w}$ records how far the
alignment front has travelled by the end of the analysis window rather than
an equilibrium radius. The warp amplitude $\psi_{\rm eff}$ depends on $e$ and $i$ in combination: among the discs
that align toward the binary plane it rises steeply with inclination
($0.08 \rightarrow 0.82$ along $e = 0$), whereas among those driven polar it
is largest just above the critical inclination and \emph{smaller} further
above it ($0.99$ at $i = 45^\circ$ against $0.32$ at $i = 60^\circ$).

\begin{table}
\centering
\caption{Disc-response summary for the twelve inclined runs. $r_{\rm w}$ is the
measured warp radius (Section~\ref{sec:theory_warp_radius}) in units of
$\ab$; $\psi_{\rm eff}$ is the warp amplitude (Appendix~\ref{apndx:detector});
$|T_\perp|/|T|$ is the out-of-plane fraction of the secular disc--binary
torque (Eq.~\eqref{eq:tperp_ratio}), averaged over $t = 500$--$1100\,\Porb$; the other quantities use the common analysis window ending at $t = 1100\,P_{\rm b}$.}
\label{tab:disc_response}
\begin{tabular}{lcccc}
\toprule
run & alignment & $r_{\rm w}$ & $\psi_{\rm eff}$ & $|T_\perp|/|T|$ \\
\midrule
$e=0.0$, $i=15^\circ$ & aligning & 6.19 & 0.08 & 1.00 \\
$e=0.3$, $i=15^\circ$ & aligning & 6.99 & 0.09 & 0.98 \\
$e=0.6$, $i=15^\circ$ & aligning & 7.29 & 0.11 & 0.99 \\
$e=0.0$, $i=30^\circ$ & aligning & 6.44 & 0.16 & 1.00 \\
$e=0.3$, $i=30^\circ$ & aligning & 8.24 & 0.18 & 1.00 \\
$e=0.6$, $i=30^\circ$ & near $i_{\rm crit}$ & 12.41 & 0.27 & 0.96 \\
$e=0.0$, $i=45^\circ$ & aligning & 9.32 & 0.30 & 1.00 \\
$e=0.3$, $i=45^\circ$ & aligning & 10.97 & 0.35 & 1.00 \\
$e=0.6$, $i=45^\circ$ & \textbf{polar} & 10.11 & 0.99 & \textbf{0.23} \\
$e=0.0$, $i=60^\circ$ & aligning & 6.99 & 0.82 & 1.00 \\
$e=0.3$, $i=60^\circ$ & near $i_{\rm crit}$ & 10.97 & 0.60 & 0.98 \\
$e=0.6$, $i=60^\circ$ & \textbf{polar} & 8.94 & 0.32 & \textbf{0.21} \\
\bottomrule
\end{tabular}
\end{table}

No disc in our grid completes a break. Two independent criteria predict that
some of them should, so the absence is itself a result. The circular-binary criterion of \citet{Nixon2013} places a
nominal tearing radius well inside the disc body, a factor $2.0$--$3.8$
interior to $r_{\rm w}$ (Eq.~\eqref{eq:r_break_nixon}), and on precession-rate grounds the two polar
discs at $e = 0.6$, with $|k|$ boosted above the coplanar rate for
$e_{\mathrm{b}} > 0.41$ (Eq.~\eqref{eq:k}), should be the most break-prone
members of the grid. Instead, all twelve inclined discs end the campaign
warping smoothly, as continuous tilt gradients rather than discontinuities. The steepest warp among
the discs that tilt toward the binary plane ($e = 0$, $i = 60^\circ$)
reached a \emph{partial} decoupling transiently and then relaxed out of it. This is the behaviour the
viscous criterion of \citet{Rabago2024} predicts at $\alpha = H/r$
(Section~\ref{sec:theory_warp_viscous}), now seen across a two-dimensional
$(e, i)$ grid rather than a single run: viscous suppression holds
both for discs that tilt toward the binary plane and for discs driven
polar, and the alignment class then sets the route by which a disc removes
its initial misalignment. Polar discs reorient their interiors coherently
onto the nearby polar orientation --- polar alignment \emph{protects} the
disc --- while discs that align toward the binary plane must transport
their full tilt viscously through the disc body, building the steepest
sustained warps and, at $i = 60^\circ$, the grid's one transient
decoupling.

Two measurements test whether that outcome is physical rather than
numerical. Both are made from the Reynolds stress carried by the resolved
velocity fluctuations --- a channel that can dominate the imposed viscosity,
as it does inside the cavity of retrograde circumbinary discs
\citep{ONeill2026} --- which we express as an effective viscosity parameter
\begin{equation}
\label{eq:alpha_reynolds}
\alpha_{\mathrm{R}}(R) \;=\;
\frac{\sum_i m_i\, \delta v_{R,i}\, \delta v_{\phi,i}}
     {\sum_i m_i\, c_{\mathrm{s},i}^2} ,
\end{equation}
where the sums run over the cells of a radial bin and over the snapshots of
the window, $\delta v$ is the departure from the mass-weighted mean
velocity of that bin and snapshot, and $c_{\mathrm{s}}$ is the local
isothermal sound speed. In the aligned discs we measure
$\alpha_{\mathrm{R}} \approx 10^{-3}$, two orders of magnitude below the
imposed $\alpha = 0.1$: velocity fluctuations on the moving mesh contribute
no significant anomalous transport, and the absence of breaking is not an
artefact of numerical diffusion. Numerical viscosity is seldom quantified in
warped-disc simulations, and where it has been it has often proved comparable
to the physical value \citep{Drewes2021}. A Reynolds decomposition cannot see
diffusion at mesh interfaces, so this bounds the resolved contribution rather
than the total effective viscosity of the scheme, which would require the
planar-spreading calibration of \citet{Drewes2021}. In the $e = 0.6$, $i =
45^\circ$ run the profile instead peaks at $|\alpha_{\mathrm{R}}| = 0.35$,
about three and a half times the imposed $\alpha$, with \emph{negative}
sign at $R \approx 8.8\,\ab$ (measured over $t = 300$--$1100\,\Porb$). The stress is coherent and radially localised: it comes from the oscillatory flow that the warp itself drives. \citet{Drewes2021} anticipated
such an effect, arguing that the local dissipation rate rises with warp
amplitude and can carry the effective viscosity above $H/r$; here it is
measured directly. Its radius differs from that of the warp, and the two separate as the run proceeds: the stress peak stays near $8.8\,\ab$ while $r_{\rm w}$ migrates outward past it (Table~\ref{tab:disc_response}), so the elevated stress marks a fixed annulus. Our strongest warps are therefore
non-linear, but the stress they generate transports the misalignment
outwards instead of breaking the disc.
\section{Orbital and accretion response of the binary}
\label{sec:orbital_evolution}

Section~\ref{sec:disc_response} established the disc-side outcome: how a disc
aligns fixes which channel of the disc--binary torque survives the secular
average. This section quantifies the binary side of that coupling. We first locate where in the disc
the torque originates (Section~\ref{sec:orbital_origin}), then follow the
orbital response through the polar-alignment transient
(Section~\ref{sec:orbital_transient}), and finally assemble the sustained
$(\adot, \edot)$ landscape across the grid
(Section~\ref{sec:orbital_maps}), whose principal result is that inclination
reverses the sign of the eccentricity forcing. Throughout, rates are
normalised by the window-mean accretion rate, i.e.\ quoted in units of
$\langle\Mdot\rangle/M_{\mathrm{b}}$, and the fixed-orbit framing applies:
these are response rates measured at prescribed $(e, i)$ for a fixed orbit (Section~\ref{sec:conclusions}).

The binary's response to those alignment states is set by how that torque divides between the two channels. The split is defined in Section~\ref{sec:methods_diagnostics}.

All discs that align toward the binary plane,
whatever their warp amplitude --- including the
partially decoupling $e=0$, $i=60^\circ$ run --- deliver their secular torque
almost entirely in the out-of-plane (precession-driving) channel,
$|T_\perp|/|T| = 0.96$--$1.00$ over $t = 500$--$1100\,\Porb$. Only two runs depart from this, and they are the two driven polar: at $e=0.6$ the fraction falls to $0.23$ ($i=45^\circ$) and $0.21$ ($i=60^\circ$). As the disc interior approaches the polar orientation, the
precession torque vanishes and the in-plane channel --- the one that drives
$\dot a$ and $\dot e$ --- takes over. The two runs launched near $i_{\rm crit}$ complete the picture: neither
reorients, and both retain $|T_\perp|/|T| \approx 1$. The out-of-plane fraction therefore drops below unity only once the polar reorientation is \emph{complete}; neither a large warp amplitude nor a launch close to the critical inclination lowers it.

\subsection{Radial origin of the torque and the precession channel}
\label{sec:orbital_origin}
\label{sec:orbital_tier2}
\label{sec:orbital_precession}

\begin{figure*}
\includegraphics[width=0.49\textwidth,height=0.46\textheight,keepaspectratio]{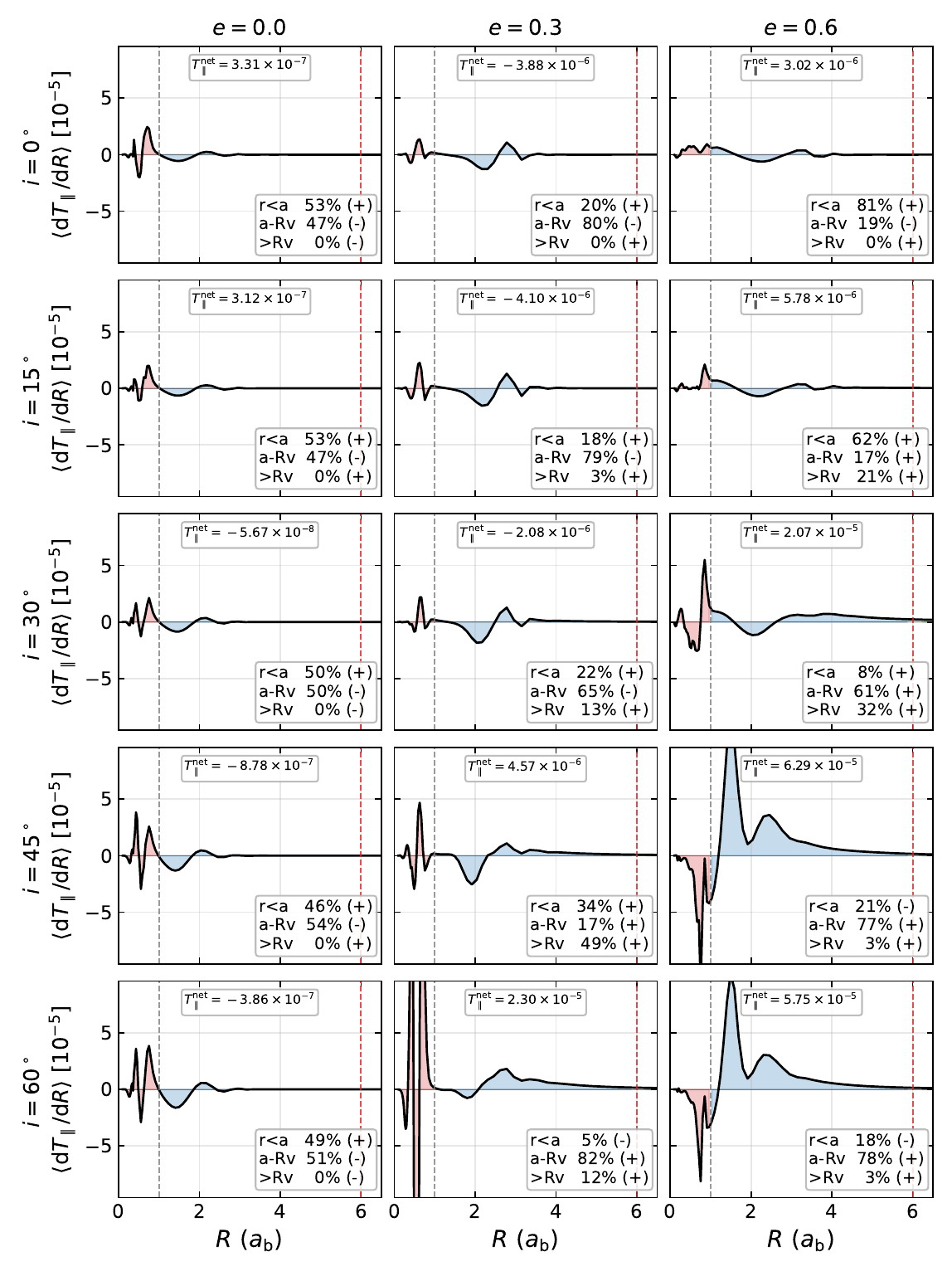}
\hfill
\includegraphics[width=0.49\textwidth,height=0.46\textheight,keepaspectratio]{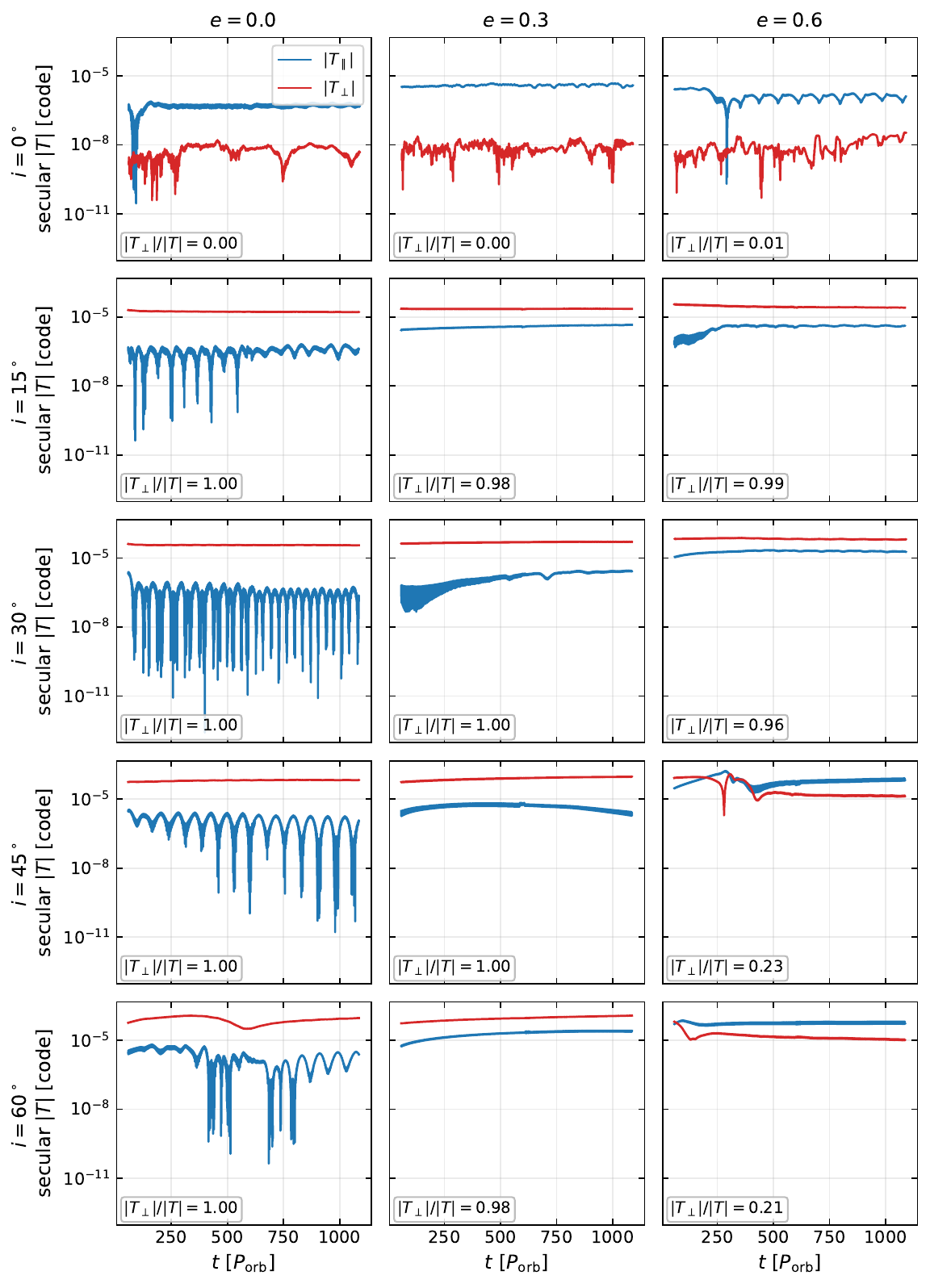}
\caption{The two decompositions of the disc--binary torque across the $(e,i)$ grid (rows: inclination; columns: eccentricity). \textit{Left:} orbit-averaged radial density of the in-plane gravitational torque, $\mathrm{d}T_\parallel/\mathrm{d}R$, from the on-the-fly $(R,z)$ accumulator, averaged over $t = 500$--$1100\,\Porb$. Shading marks the three radial zones whose shares, computed over the full disc, are annotated in each panel: $r < \ab$ (red; cavity and circumsingle region), $\ab < r < \Rvisc$ (blue; inner circumbinary disc) and $r > \Rvisc$ (green; outer disc). A contribution's sign is set by whether the curve lies above or below zero; the radial axis is truncated at $6.5\,\ab$. \textit{Right:} the secular in-plane ($|T_\parallel|$, blue) and out-of-plane ($|T_\perp|$, red) components versus time, with the mean fraction $|T_\perp|/|T|$ over the same window (Eq.~\eqref{eq:tperp_ratio}) annotated. Discs that align toward the binary plane deliver essentially all of their secular torque in the out-of-plane channel; only the two polar discs bring the fraction down, to $0.23$ and $0.21$.}
\label{fig:torque_regional}
\end{figure*}

Figure~\ref{fig:torque_regional} shows the orbit-averaged radial density
$\mathrm{d}T_\parallel/\mathrm{d}R$ of the in-plane gravitational torque on
the binary (the $\parallel$ component of Eq.~\eqref{eq:tsplit}), from the on-the-fly $(R,z)$ diagnostic
(Section~\ref{sec:methods_diagnostics}). Two structural features stand out. First, the cavity and stream region ($r < \ab$) and the inner circumbinary disc exert torques of opposite sign in twelve of the fifteen runs, so that the net secular torque is their difference; in the five $e = 0$ runs they nearly cancel, each exceeding the net by factors of about 6 to more than 20. In the other three runs ($e = 0.6$ at $i = 15^\circ$ and $30^\circ$, and $e = 0.3$ at $i = 45^\circ$) all zones push the same way. This cancellation structure is the underlying
reason net CBD torques are notoriously sensitive to resolution and numerical
scheme \citep{Munoz2019, Moody2019}: modest changes in the cavity-edge torque
density can flip the sign of the residual. Second, the torque is
inner-dominated: it accumulates almost entirely within $r \lesssim 8$--$10\,\ab$,
even though the disc extends to $50\,\ab$ and its mass and angular momentum
are dominated by large radii --- the same localisation \citet{Roedig2012}
found in 3D for massive self-gravitating discs, where the gravitational torque
is carried by overdensities at the disc edge. The binary's orbital response is therefore set
at the cavity edge, by gas whose alignment state is exactly the quantity that Section~\ref{sec:branches} shows the alignment to control.

A snapshot-based reconstruction of the same force (Section~\ref{sec:methods_diagnostics}) gives the phase-robust ratio
$T_{\mathrm{3D}}/T_{\mathrm{2D}}$ between the full 3D torque and the torque
from a $z=0$ projection of the same snapshot, which is $\approx 1$ for
aligned discs: gravitational $z$-dilution is small at $i = 0^\circ$,
consistent with \citet{Moody2019}, whose $\sim 23\%$ 3D--2D difference lies
in the advective channel rather than in gravity.

For every disc that aligns toward the binary plane the secular torque is
delivered almost entirely in the out-of-plane channel
($|T_\perp|/|T| \geq 0.97$; Section~\ref{sec:phasemap}). The corresponding
effect is purely three-dimensional: were the orbit live, the disc torque
would reorient (precess) the binary orbital plane on the timescale

\begin{equation}
\label{eq:tau_prec}
\tau_{\mathrm{prec}} \;=\; \frac{\ell_{\mathrm{b}}}{|\langle \vect{T}_\perp \rangle|} ,
\qquad \ell_{\mathrm{b}} = \sqrt{1 - e_{\mathrm{b}}^2} ,
\end{equation}
where $\ell_{\mathrm{b}}$ is the binary's specific orbital angular momentum.
The order of operations matters: we take $|\langle \vect{T}_\perp \rangle|$
and not $\langle |\vect{T}_\perp| \rangle$. The torque direction swings over
the binary orbit, so averaging the vector first lets those oscillations
cancel and leaves the secular part that actually drives precession.
Evaluated over $t = 500$--$1100\,\Porb$ this gives $\tau_{\mathrm{prec}} \approx (1.5$--$9.5) \times 10^3\,\Porb$ for the discs that tilt toward the plane, generally shortening with $i$ and $e$, and $8.5 \times 10^3$ and $1.1 \times 10^4\,\Porb$ for the two polar discs, while for $i = 0^\circ$ the perpendicular torque vanishes by symmetry and $\tau_{\mathrm{prec}}$ exceeds $10^7\,\Porb$. Because our
binary is fixed, what the diagnostics record is the disc reorienting relative
to the binary; $\tau_{\mathrm{prec}}$ is the timescale on which a live binary
would respond (Section~\ref{sec:methods_binary}). In the two polar runs this channel shuts down, because the polar configuration is a stationary
point of the secular dynamics about which a nearly polar disc only
oscillates \citep{Lubow2018}, and the in-plane channel --- the one that
drives $\adot$ and $\edot$ --- inherits the torque. What that reorganisation
looks like in time is the subject of the next subsection.

\subsection{The polar-alignment transient}
\label{sec:orbital_transient}

At $e = 0.6$, $i = 45^\circ$ the reorientation of Section~\ref{sec:branches} is a single dynamical event. The specific angular momentum
delivered by accretion turns retrograde first, at $t \approx 210\,\Porb$ (see Figure \ref{fig:delta_grid});
the cavity then shrinks from $2.8$ to $1.3\,\ab$; the accretion rate rises to $\approx 2.7$ times its secular mean at $t \approx 255\,\Porb$;
and at $t \approx 288\,\Porb$, once the inner-disc tilt crosses
$\delta \approx 80^\circ$, the out-of-plane torque fraction drops to its polar value and $\adot/\ab$ reaches its deepest excursion,
$(\adot/\ab)/\langle\Mdot\rangle \approx -115$. The out-of-plane fraction does not lag the tilt by a fixed interval; it
changes only once the tilt crosses a threshold. Through $\sim\!150$ orbits
of smooth tilt growth the fraction stays at unity, and it falls only as the
interior nears the polar configuration. The gravitational torque does the work --- it peaks exactly as the fraction drops, while the direct accretion torque is $\sim\!100$ times
smaller throughout --- so the angular-momentum reversal at $t \approx
210\,\Porb$ leads the event as a tracer rather than causing it. It
is worth recording what that reversal measures, since accretion is one of
the three channels of Section~\ref{sec:theory_orbit}. We track $\vect{l} =
\Delta\vect{L}_{\rm acc}/\Delta M$, the change in the binary's
\emph{relative} specific angular momentum per unit mass accreted, in units of $\sqrt{G M_{\mathrm{b}}\ab}$. It differs from the scalar $\ell_0 =
\dot{J}_{\mathrm{b}}/\dot{M}_{\mathrm{b}}$ of the 2D coevolution literature:
recovering $\ell_0$ requires a reduced-mass term that we do not track and
that dominates at $q = 1$, so we make no comparison with published values of
it. Accumulating the same quantity sink-step by sink-step and from interval-
aggregated momenta agrees to $0.2$--$6$ per cent across the coplanar runs,
so the coarse-grained bookkeeping does not bias it. Its sustained magnitude is small
around circular binaries ($|\vect{l}| \leq 0.23$ at every inclination) and
strongly retrograde for eccentric ones even when the disc is coplanar,
reaching $l_{\mathrm{acc},\parallel} = -1.88$ at $e = 0.6$, $i = 0$ and
weakening with inclination ($-1.50$, $-0.29$, $-0.24$ at $i = 15^\circ$,
$30^\circ$, $45^\circ$). These are rates per unit accreted mass, and because
the mass accreted per orbit is small the channel does not drive the orbital
evolution: its value here is as the earliest signature of the transition.

The second polar disc, at $i = 60^\circ$, repeats the sequence in
compressed form and reveals what sets its character. Beginning well above
the critical inclination, it is retrograde in delivery from the start, with
no prograde phase and no sign flip; its cavity shrinks by $t \approx 60\,\Porb$; and instead of rising its accretion rate
\emph{dips} to $\approx 0.3$ of the secular mean. Its out-of-plane fraction drops over $t \approx 75$--$115\,\Porb$, beginning at the same
$\delta \approx 80^\circ$ threshold. The distance above the critical
inclination therefore sets both how fast the transition happens and what it looks like: an onset from just above the boundary passes through a
prograde phase and a rise in accretion, while one from well above is born
retrograde and suppresses accretion instead. Both converge on the same
polar end state, whose sustained orbital shrinking is an order of magnitude
beyond the aligned-disc values (Section~\ref{sec:orbital_maps}).
Appendix~\ref{apndx:transients} presents the full diagnostic sequence.

For a live binary these are simultaneous electromagnetic and
orbital-element signatures of a single internal disc transition; we return
to the observational implications in Section~\ref{sec:discussion}.

\subsection{Sustained response: inclination circularises the binary}
\label{sec:orbital_maps}
\label{sec:orbital_results_tier1}
\label{sec:orbital_e0}

\begin{figure*}
\includegraphics[width=\textwidth]{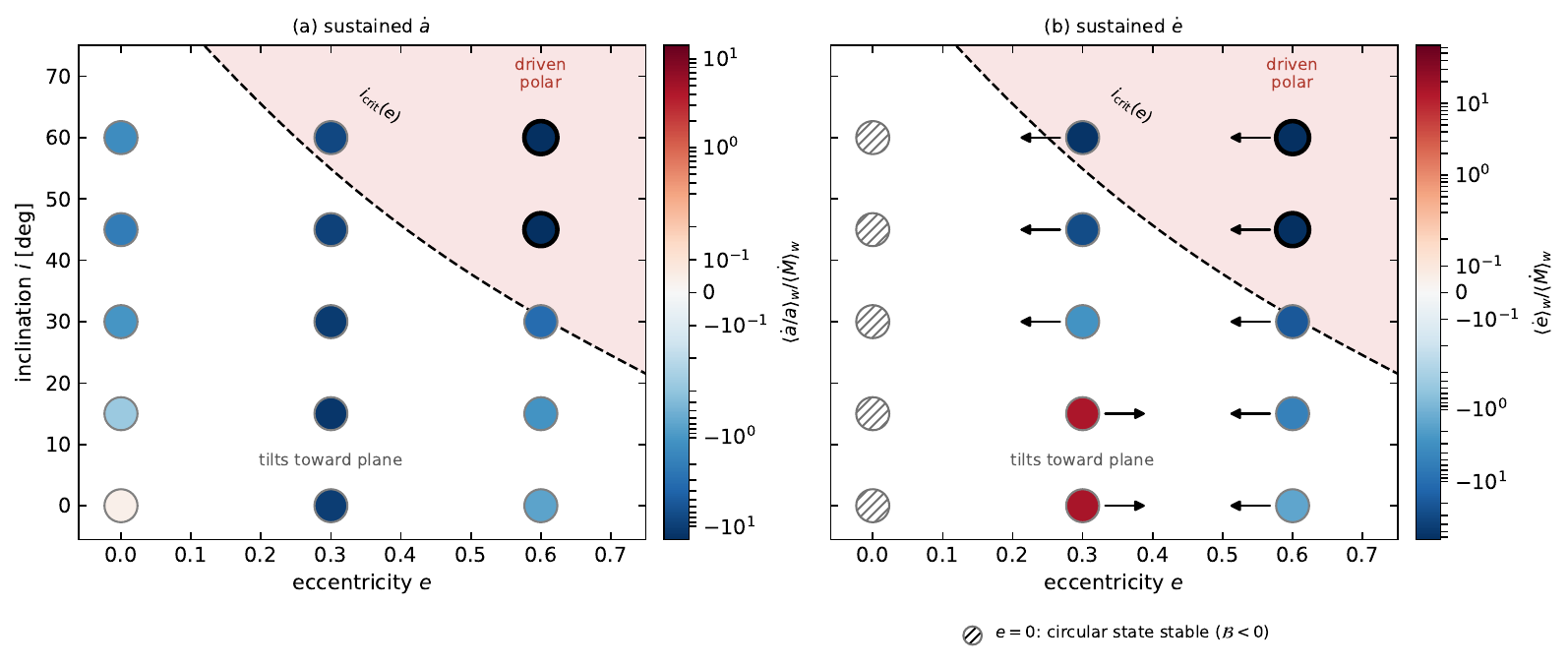}
\caption{Sustained orbital response across the $(e,i)$ grid: window means of (a) $\adot/\ab$ and (b) $\edot$, normalised by $\langle\Mdot\rangle$, over $t = 500$--$1100\,\Porb$ (symmetric-log colour scale; red = expansion/pumping, blue = shrinking/damping, sign convention of \citealt{Siwek2023a}). Colour ranges are set by the largest value outside the two polar runs, whose cells (thick black edges) saturate; their values are given in the text. The dashed curve is $i_{\rm crit}(e)$, with the polar side shaded. In panel (b) the $e = 0$ symbols are hatched, as $\edot$ is formally $0/0$ there; the bracket $\mathcal{B}$ (Eq.~\eqref{eq:bracket}) is negative at every inclination, so the circular state is stable. The fixed-length arrows in panel (b) give the sign of $\edot$ (right: pumping; left: damping) at a fixed orbit. At $e = 0.3$ the pumping at $i \leq 15^\circ$ reverses to damping for $i \geq 30^\circ$.}
\label{fig:orbital_maps}
\end{figure*}

Figure~\ref{fig:orbital_maps} assembles the sustained orbital response ---
the window-mean $\adot/\ab$ and $\edot$ over the common window $t = 500$--$1100\,\Porb$, which begins after the turns to polar of Section~\ref{sec:orbital_transient} --- across the $(e,i)$ grid. Quoted ranges are $1\sigma$ uncertainties of these window means (the 16th to 84th percentiles over 500 resamplings of 50-orbit blocks of each series; halving or doubling the block length changes no conclusion).

\textit{The circular attractor is stable at every inclination, and the
damping is stronger at high inclination.} For the nominally circular runs, $\edot$ is formally $0/0$ and the
bracket $\mathcal{B}$ (Eq.~\eqref{eq:bracket}) is the well-defined stability
diagnostic. Its secular mean is negative in all five $e = 0$ runs:
$\langle\mathcal{B}\rangle/\langle\Mdot\rangle = -9.3,\,-9.7,\,-10.9,\,-13.8,\,-21.3 \times 10^{-3}$ at $i = 0^\circ, 15^\circ, 30^\circ, 45^\circ, 60^\circ$, measured over $t = 500$--$1100\,\Porb$, all negative well beyond their uncertainties. The damping strengthens with inclination, and the differences between neighbouring inclinations exceed the uncertainties from $i = 30^\circ$ upward. These magnitudes grow through the integration, by about 40 per cent between the full run and $t = 500$--$1100\,\Porb$ and by another 25 per cent in the last 250 orbits, so we quote them with the window stated; what does not change with the
window is the sign at every inclination and the excess at high inclination,
and it is those the claim rests on.
For near-circular binaries, residual eccentricity is damped at every
inclination we simulate --- the 3D,
misaligned extension of the circular fixed point found in 2D
\citep{Munoz2019, DOrazio2021, Zrake2021} --- and misalignment strengthens the damping.

\textit{Inclination reverses the eccentricity forcing at $e = 0.3$.} In the
coplanar row our signs reproduce the pattern found in earlier work
\citep{Roedig2011, Siwek2023a, DOrazio2021, Zrake2021}: $\edot > 0$ at $e = 0.3$, $\edot < 0$ at
$e = 0.6$, implying the familiar eccentric attractor at
$e_{\mathrm{eq}} \approx 0.4$--$0.5$. At $e = 0.3$, however, the pumping weakens as the inclination grows and flips sign between $i = 15^\circ$ and $30^\circ$:
$\langle\edot\rangle/\langle\Mdot\rangle \approx
+16,\,+15,\,-2.6,\,-28,\,-54$ at $i = 0^\circ$--$60^\circ$. The reversal lies far outside the uncertainties: $+14.6\,[+14.3, +14.9]$ at $i = 15^\circ$ against $-2.6\,[-3.1, -2.1]$ at $i = 30^\circ$. The reversal occurs deep inside the aligning outcome ($i_{\rm crit}(0.3) = 54.9^\circ$, so the $i = 30^\circ$ and $45^\circ$ discs are ordinary warped discs aligning toward the binary plane), so inclination alone causes it, without any polar reorientation. At $e = 0.6$ the response is damping at every inclination, and steepens
sharply in the two discs driven polar
($\langle\edot\rangle/\langle\Mdot\rangle \approx -235$ and $-302$ at $i =
45^\circ, 60^\circ$). Taken together with the stronger high-inclination damping at $e = 0$, the two measurements agree: \emph{in 3D, inclination
favours circularisation across the range we sample}. Beyond our grid, at the
retrograde extreme, discs can instead drive eccentric inspirals, depending on
the rotation sense of their minidiscs \citep{ONeill2026}. The 2D eccentric attractor exists at
low inclination but is gone --- moved below $e = 0.3$ or erased --- by
$i \approx 30^\circ$. Long-term secular models built on the 2D suites
\citep{Valli2024} therefore describe the coplanar regime; for misaligned discs
the attractor they integrate toward is absent, although how far a binary
actually moves in either regime depends on a disc mass budget that our
fixed-orbit runs do not address. Since gas-disc eccentricity pumping is a
leading channel for building eccentric (massive black-hole) binaries, an
inclination-dependent $e_{\mathrm{eq}}$ landscape feeds directly into
predictions for binary populations and their gravitational-wave residuals;
we develop this in Section~\ref{sec:discussion}.

\textit{Semi-major-axis response.} The $\adot$ panel shows the complementary
structure: at low inclination the $e = 0.3$ column shrinks most rapidly
(again matching the 2D sign structure), and this weakens steadily toward
high inclination, while the two polar runs
sit far off the quiet scale at $(\adot/\ab)/\langle\Mdot\rangle \approx
-129$ and $-145$ --- the sustained $10$--$20\times$ enhancement in orbital
shrinking established
in
Section~\ref{sec:orbital_transient}, still deepening at the end of the
chains.

The instantaneous rates oscillate at $\pm
100$--$200$ on the orbital timescale, so all values quoted are secular
window means. The reversal holds for any choice of window: recomputing on
windows from the last $150$ to the full $1050\,\Porb$ changes no sign and no
ordering, and does not move the location of the reversal. Amplitudes are
less stable than signs --- the damping at $e = 0.3$, $i = 30^\circ$ weakens
monotonically as the window weights later times --- so we quote the reversal
as a sign change and not as a magnitude. The two polar runs behave differently, and informatively: at $i = 60^\circ$,
which turns polar at $t \approx 115\,\Porb$, the full-integration $\adot$
agrees with the value over $t = 500$--$1100\,\Porb$ to $4$ per cent ($-139.4$ against $-144.9$), whereas at $i = 45^\circ$, which does not turn until $t \approx
290\,\Porb$, averaging over the full integration reduces it by about a third
($-84.0$ against $-129.6$). The enhanced shrinking is a property of the
polar state itself, so averaging across the transition dilutes it, more so the longer the disc spent before turning polar: that phase fills 23 per cent of the full run at $i = 45^\circ$, but only 6 per cent at $i = 60^\circ$. Once both discs are polar, the window no longer matters: over $t = 850$--$1100\,\Porb$ the two rates are $-145.2$ and $-129.3$, within 0.2 per cent of their values over $t = 500$--$1100\,\Porb$.  The circular coplanar run has the smallest rate in the grid, and its sign depends on the window: $\adot/\ab \approx +0.04$ per accretion time over $t = 500$--$1100\,\Porb$, $+0.10$ over $t = 850$--$1100\,\Porb$ and $-0.05$ over the full run.

The binary columns of Table~\ref{tab:timescales} show how the alignment state
sets the binary's response. Around discs that tilt toward the plane, the
out-of-plane torque dominates: it mainly precesses the binary's orbital
plane about the disc's angular momentum, with periods of $(0.6$--$2.4) \times 10^4\,\Porb$ for the simulated disc, and tilts it more
slowly toward the disc. Compared through these timescales, whose ratios do not depend on the disc mass, the inclination then changes about as fast as the eccentricity and faster than the semi-major axis ($t_e$ lies within a factor of 2.5 of $t_i$ in four of the six eccentric runs, and $t_a$ exceeds $t_i$ by factors of 4--110). Around polar discs the
torque passes to the in-plane channel: the binary tilts slowly away from the
disc, toward a perpendicular orientation, while $e$ and $a$ change 14--20 and about five times faster than $i$, and per unit time the binary shrinks 70--90 times faster and circularises 65--75 times faster than around a coplanar disc
of the same eccentricity. Within our grid, polar discs are therefore by far
the most efficient drivers of orbital decay and circularisation, although
with two polar runs, both at $e = 0.6$, we cannot yet say how this
efficiency depends on eccentricity. The disc reaches its alignment state
first: for the simulated disc mass, polar discs align two to three orders of
magnitude faster than the binary plane precesses, and discs that tilt toward
the plane a few times faster. Because the binary's timescales scale with the
accretion time, $(0.7$--$4.7) \times 10^5\,\Porb$ here, the disc settles
first at any lower accretion rate.

\subsection{Accretion variability as an alignment-state diagnostic}
\label{sec:disc_observability}

\begin{figure*}
\includegraphics[width=\textwidth]{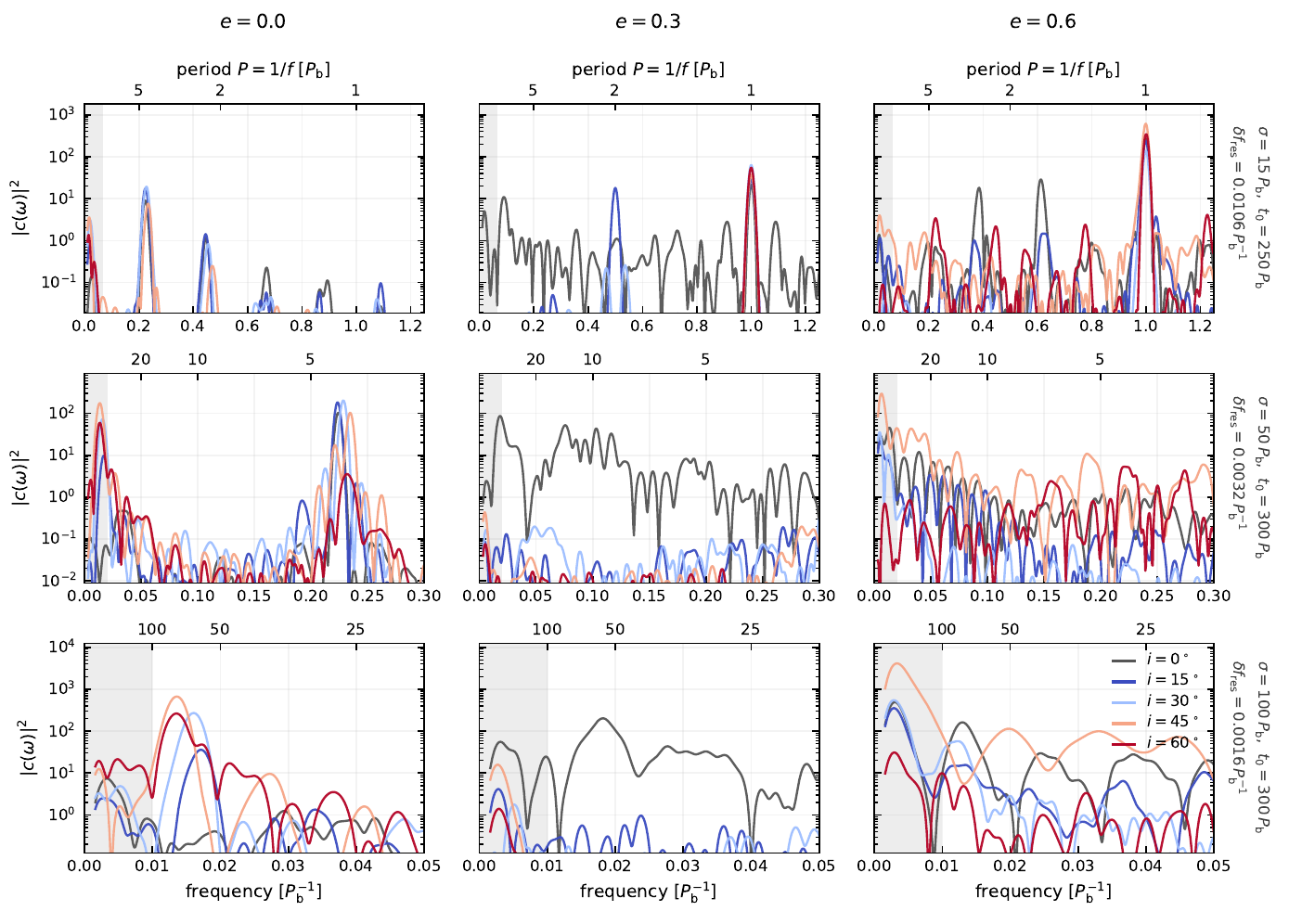}
\caption{Accretion variability across the grid: Gabor periodograms $|c(\omega)|^2$ of $\Mdot/\langle\Mdot\rangle$ (columns: $e$; colours: $i$) for Gaussian windows of width $\sigma = 15$, $50$ and $100\,\Porb$ (rows, top to bottom; centred at $t_0 = 250$, $300$ and $300\,\Porb$). Each row shows only the frequencies its window resolves ($\delta f = 1/(2\pi\sigma)$); the grey band marks $f < 1/\sigma$, where fewer than one cycle fits in the window. The top axes give the period $P = 1/f$. At $e = 0$ the lump ($P \approx 4.5\,\Porb$) weakens with inclination and vanishes at $i = 60^\circ$, while a $60$--$75\,\Porb$ peak appears in every inclined run, strongest at $i = 45^\circ$; at $e > 0$ the orbital peak persists at every inclination, including both polar runs and the $e = 0.3$, $i = 15^\circ$ run, where a sub-harmonic at $P = 2\,\Porb$ is also present.}
\label{fig:periodograms}
\end{figure*}

On observational timescales the binary's orbital elements are frozen;
the accessible signature of the disc--binary coupling is the accretion
rate, whose variability is already known to encode the disc's state: in
coplanar 2D discs the ratio of the two components' accretion rates varies
on the apsidal precession period of the circumbinary disc
\citep{DeLaurentiis2026}. Figure~\ref{fig:periodograms} condenses $\Mdot(t)$ across the grid into
Gabor periodograms of the fractional accretion rate
$\Mdot/\langle\Mdot\rangle$ --- Fourier transforms
of the series weighted by a Gaussian window of width $\sigma$ centred on a
chosen epoch, so that power is localised in time as well as in frequency ---
at three window widths: $\sigma = 15\,\Porb$, matching the analysis of
\citet{Duffell2024}; an intermediate $\sigma = 50\,\Porb$; and $\sigma =
100\,\Porb$. Only the widest window resolves periods of $50$--$100\,\Porb$. Each row shows only the
frequency range its window legitimately resolves (see caption). The choice of estimator is not critical: a Lomb--Scargle periodogram --- a
least-squares sinusoid fit to the whole series, with no time window ---
recovers the lump, orbital-period and $50$--$100\,\Porb$ peak frequencies to better than one per cent, and finds no $50$--$100\,\Porb$ peak in the coplanar $e = 0$ control. The two estimators disagree only for the $e = 0.6$, $i = 45^\circ$ run, whose series contains the one-off transient of Section~\ref{sec:orbital_transient}: a global periodogram redistributes that single event into spurious long-period power. Such non-stationarity is the reason for the time-localised Gabor form, together with its direct comparability with \citet{Duffell2024}.

The coplanar ($i = 0^\circ$) curves reproduce the canonical anatomy of circumbinary
variability: at $e = 0$ the cavity-edge overdensity (the ``lump'') modulates
accretion at $\approx 4.5\,\Porb$, giving the comb at $f \approx
0.22\,\Porb^{-1}$ and its harmonics \citep{MacFadyen2008, ShiKrolik2012, Farris2014, Duffell2024}, while eccentric binaries are forced at the orbital
period \citep{Munoz2016, Duffell2024}. Across the grid these features differ in how well they survive misalignment:

\begin{enumerate}

\item \textit{The $1\,\Porb$ eccentric-forcing peak is the most robust
observable.} It is present at every inclination in every $e > 0$ run, and it survives
polar alignment: both $e = 0.6$ polar runs retain it as their dominant short-
period peak in post-transient windows ($|c(\omega)|^2 \approx 3 \times 10^2$
at $t_0 = 450\,\Porb$ for both). It is also the feature \citet{Duffell2024}
identify as least sensitive to the sink prescription. A dominant $1\,\Porb$ peak indicates an eccentric binary
regardless of alignment state. Its phase may still carry the alignment: around
$e_{\mathrm{b}} = 0.5$ binaries, the dominant accretion pulse moves from before
to after periastron as the misalignment grows from $60^\circ$ to $120^\circ$
\citep{Yang2026Pulsed}. Our periodograms measure power rather than phase, so the two
diagnostics are complementary. At $e = 0.3$, $i = 15^\circ$ an additional peak exists at $2\,\Porb$ ($f \approx 0.5\,\Porb^{-1}$; top row of Fig.~\ref{fig:periodograms}), a sub-harmonic of the orbital forcing that stays within a factor $1.7$ of the orbital power throughout the run. Successive pericentre bursts alternate in strength, a strong one at
$\approx 2.5$ times the mean rate followed by a weak one at $\approx 1.5$
times, so the pattern repeats only every second orbit. Both black holes
accrete strongly in the same orbits, so the accretion event as a whole
alternates. No other run shows it at comparable strength: only the
$e = 0.3$, $i = 30^\circ$ run develops a sustained version, about ten
times weaker, in its last $\sim 250$ orbits, and no $e = 0$ or $e = 0.6$
run shows one. Because the coplanar $e = 0.3$ run shows no such alternation while the $i = 15^\circ$ run shows it throughout, an alternating, every-second-orbit burst pattern is a candidate signature of a moderately eccentric binary slightly inclined to its disc; establishing it as a diagnostic needs a finer sampling of $e$ and $i$ around this run.

\item \textit{The lump is an inner-disc alignment-state diagnostic.} This is
the time-domain counterpart to the structural diagnostic of
\citet{Franchini2019}, in which the cavity radius itself carries the
misalignment. The $4.5\,\Porb$ comb survives misalignment through $i =
45^\circ$ at
$e = 0$ but is destroyed in the partially decoupling $i = 60^\circ$
disc during its misaligned epoch. In the window shown that run has no short-period peak at all: every other
run peaks at short period ($4.3$--$4.5\,\Porb$ at $e = 0$, $1\,\Porb$ at $e
> 0$), whereas its spectrum rises monotonically toward long periods,
carrying $1.0$--$2.9$ dex less short-period power than any other run. Its long-period variability instead follows its inner disc, which during the decoupling precesses independently of the outer disc (Section~\ref{sec:branches}): between $t \approx 300$ and $700\,\Porb$ the accretion rate tracks the inner disc's node, a 24 per cent modulation at its $\approx 435\,\Porb$ precession period. It is the only run whose inner disc completes a precession cycle within the simulated time.
The absence of short-period power is confined to that epoch. As
the window is slid to later epochs the lump returns at the period shared by
the other $e = 0$ runs and grows monotonically by two orders of magnitude,
tracking the alignment of the inner disc as its tilt falls below $\delta
\approx 25^\circ$. It recovers while the outer disc still stands at $\Theta
\approx 74^\circ$ against an inner disc at $\delta \approx 18^\circ$. The lump thus tracks the \emph{inner}-disc
alignment state, insensitive to the state of the disc beyond the decoupling
radius; the recovery of the lump and the relaxation of the decoupling described in Section~\ref{sec:warp_no_break} are the same turnover.

\item \textit{Inclination quenches the broad eccentric-coplanar power.}
At $e = 0.3$ the coplanar run carries broad $5$--$100\,\Porb$ power
(apsidal- and lump-related); inclination suppresses it monotonically, by up
to $\approx 3.5$ dex at $i = 60^\circ$.

\end{enumerate}

A weaker modulation at $50$--$100\,\Porb$ appears in many runs and does not
trace the alignment state. It is present in every inclined $e = 0$ run,
strongest at $i = 45^\circ$, and in several eccentric runs, including both
coplanar ones; the coplanar circular control shows none. In the coplanar eccentric runs it is a harmonic of the prograde precession of the eccentric cavity edge, at half ($e = 0.6$) and a third ($e = 0.3$) of its $167$--$174\,\Porb$ period; the fundamental appears as the alternation of the accretion between the two black holes (Appendix~\ref{apndx:gallery}), as in the SPH simulations of \citet{Dunhill2015}. In the inclined
$e = 0$ runs its source is an oscillation in the orientation of the cavity edge,
by $3.5$--$3.9^\circ$ in node and $0.6$--$1.7^\circ$ in tilt at
$r = 1.5$--$2\,\ab$, which travels across the inner disc and modulates the
accretion rate by 4--20 per cent. The inner disc's secular precession is
much slower, with periods of $\approx 435\,\Porb$ in the decoupling run and
$(2$--$5) \times 10^4\,\Porb$ in the other inclined $e = 0$ runs, and the
strongest example of the oscillation is the $e = 0$, $i = 45^\circ$ disc, which
never approaches decoupling. The period of the oscillation matches, to within
5--8 per cent, the local nodal precession period of a ring at
$r \approx 2\,\ab$ inclined at the inner disc's measured tilt. That radius also lies at the outer edge of our super-Lagrangian refinement region (Appendix~\ref{apndx:refinement}), but the oscillation is not tied to it: restarting this run at $t = 300\,\Porb$ with that edge moved to $1.5$ or $3\,\ab$ leaves its amplitude, radial profile, accretion modulation and implied radius unchanged, with a period about 12 per cent longer in both restarts, comparable to the change of the run's own period from $72$ to $78\,\Porb$ between $t = 500$--$700$ and $700$--$900\,\Porb$.

Taken together, the three diagnostics classify the system from the
periodogram alone. A lump comb without a $1\,\Porb$ peak indicates a near-
circular, near-coplanar binary; a dominant $1\,\Porb$ peak indicates an
eccentric binary in any alignment state; and the loss of short-period power, together with an accretion modulation at the inner disc's own precession period, indicates a strongly warped disc approaching decoupling, a state for which we are not aware of an earlier time-domain signature in a circumbinary disc (for tearing discs around spinning black holes, see \citealt{RajNixon2021}). The subsequent return of the lump signals an aligned interior below a still-decoupled exterior.

The polar-alignment transient, by contrast, leaves no periodogram feature. It
happens once, so no window recovers it as a periodicity, and its observable
is the light curve itself: an accretion peak if the disc starts just above
the critical inclination, a suppression if it starts well above it,
simultaneous with the orbital-element excursion of
Section~\ref{sec:orbital_transient}. Catching it therefore requires
monitoring rather than a periodicity search, and
Section~\ref{sec:disc_implications} sets its duration against survey
baselines.

\begin{figure}
\includegraphics[width=\columnwidth]{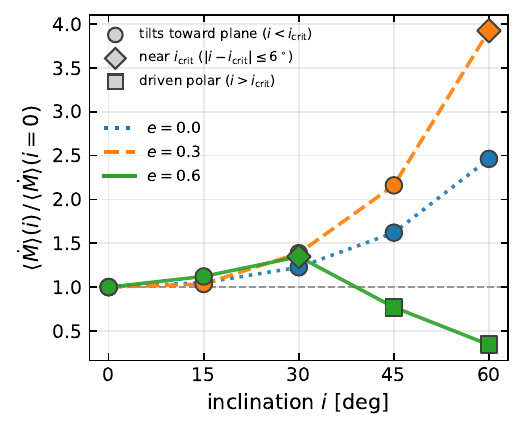}
\caption{Mean accretion rate versus inclination over $t = 100$--$1100\,\Porb$, normalised for each eccentricity by its coplanar run, $\langle\Mdot\rangle(i)/\langle\Mdot\rangle(i=0)$, since the absolute rate differs by a factor of $\sim 3$ between eccentricities. Eccentricity is given by colour and linestyle; symbol shape gives the predicted alignment class (circle: $i < i_{\rm crit}$; diamond: within $6^\circ$ of $i_{\rm crit}$; square: $i > i_{\rm crit}$, driven polar). Discs that align toward the binary plane brighten monotonically with misalignment, by factors of $2.5$ and $3.9$ at $e = 0$ and $0.3$; the polar discs at $e = 0.6$ are suppressed, to $0.77$ ($i = 45^\circ$) and $0.34$ ($i = 60^\circ$).}
\label{fig:mdot_normalised}
\end{figure}

The mean accretion level carries the alignment signature as well as its
variability, as Fig.~\ref{fig:mdot_normalised} shows. Because the
absolute accretion rate varies by a factor of three between eccentricities, the comparison has to be made separately for each eccentricity, against its own coplanar run; compared globally, that threefold spread masks the trend
entirely. So normalised, the mean rate over $t = 100$--$1100\,\Porb$ rises monotonically with
misalignment for the two lower eccentricities, reaching
$\langle\Mdot\rangle(i)/\langle\Mdot\rangle(0) = 2.46$ at $e = 0$ and $3.93$ at $e = 0.3$ by $i = 60^\circ$ (the full sequences are $1.05,\,1.23,\,1.62,\,2.46$ and $1.04,\,1.39,\,2.16,\,3.93$ from $i = 15^\circ$). A disc that is
misaligned accretes substantially faster than the same disc in the plane.
This is the circumbinary counterpart of the effect \citet{Lodato2006}
identified for warped discs around spinning black holes, where the
dissipation associated with twisting the disc raises the accretion rate and
renders misaligned systems more luminous than flat ones.

The polar discs depart from that pattern, and only a grid that contains both
alignment outcomes can show it. At $e = 0.6$ the enhancement
does not appear: the two runs below the critical inclination are modestly brightened ($1.12$ and $1.35$), but both discs driven polar are
\emph{suppressed} --- to $0.77$ at $i = 45^\circ$ and $0.34$ at
$i = 60^\circ$. Polar alignment removes
the sustained twisting that a disc aligning through the plane undergoes, and
with it the dissipative brightening. This happens irrespective of warp
amplitude: the $i = 45^\circ$ polar disc carries the largest warp of the two
($\psi_{\rm eff} = 0.99$) and is suppressed all the same.
Among the discs that align toward the binary plane we cannot separate the
roles of warp amplitude and inclination: the two co-vary almost perfectly
across our ten aligning runs, and they rank the normalised accretion rate
almost equally well, with Spearman rank correlations --- a measure of
monotonic association between rankings rather than of linear fit --- of
$\rho = 0.98$ and $0.97$ respectively. What the data do support is a
dependence on which alignment the disc adopts. A misaligned circumbinary system
should therefore be at its most luminous while it is still misaligned and aligning toward the plane, that is, during its approach to coplanar alignment. Both discs driven polar are instead suppressed. With only two polar runs,
both at $e = 0.6$, we read that as a qualitative difference between the two alignment outcomes; a calibrated prediction would need more polar runs.

One caveat tempers the mapping to observed light curves: our observable
is the accretion rate onto the sinks, not a luminosity, and the two need not
track each other --- a binary whose accretion is suppressed can remain
luminous and variable in the optical \citep{Tiede2026}. The conversion
passes through the minidiscs and the radiative efficiency, which can
buffer and low-pass filter the fastest variability, so the
\emph{frequencies} of the features above are more robust predictions
than their amplitudes.

\section{Discussion}
\label{sec:discussion}

\subsection{3D versus 2D: thermodynamic regime and the 3D torque channels}
\label{sec:disc_3d_vs_2d}

The thermodynamic regime of our disc (locally-isothermal $\alpha$-disc,
$\alpha = H/r = 0.1$) is the direct 3D analogue of the \citet{Munoz2019},
\citet{DOrazio2021}, and \citet{Siwek2023a} 2D simulations. For binary
accretion, the locally isothermal approximation is least consequential at
this thickness: when the energy equation is evolved self-consistently in 2D,
accretion is suppressed only in thin, high-Mach discs, while runs near our
$H/r$ equilibrate rapidly \citep{Tiede2026}. In this regime
we expect the cavity-region positive torque to be competitive, in contrast to the
gravito-turbulent \citet{Bourne2024} regime where the outer-disc torque dominates.
The $\sim 23\%$ advective-torque suppression found by \citet{Moody2019} in 3D
relative to 2D, for the same thermodynamic regime, motivates our snapshot-based
gravitational reconstruction
(Section~\ref{sec:orbital_origin}). That reconstruction returns $T_{\mathrm{3D}}/T_{\mathrm{2D}} \approx 1$ for
the aligned discs: the gravitational torque is not diluted by vertical
structure, so the suppression they report resides in the advective channel,
carried by vertical angular-momentum flux. Quantifying that channel directly requires snapshot output at sub-orbital
cadence, so that the velocity field can be phase-averaged over the binary
orbit; we defer this to future work.

The binary precession channel (Section~\ref{sec:orbital_precession}) is the most
securely identified purely-3D result: it is zero by construction in 2D axisymmetric
simulations and arises entirely from the out-of-plane tidal torque of the warped
disc on the binary. This channel matters for SMBH binaries in AGN, where gas is delivered by
mergers at an orientation uncorrelated with the binary plane, so
misalignment is the generic expectation \citep[e.g.][]{King2006, Nixon2011, Goicovic2016, Goicovic2017}.

\subsection{Comparison with disc-breaking criteria}
\label{sec:disc_regime}

None of our discs breaks, in agreement with the $\alpha = H/r$ run of
\citet{Rabago2024}, which warps but does not break
(Section~\ref{sec:theory_warp_viscous}). We attribute this primarily to the
viscosity. A disc breaks only where the stresses that communicate the warp
are weak compared with the differential precession: in the diffusive regime,
breaks form where strong warping lowers the effective viscosity non-linearly \citep{Ogilvie1999, Lodato2010, NixonKing2012}, and in the wave-like regime tearing appears once the
total (physical plus numerical) viscosity is small and the misalignment large
\citep{Nealon2015, Drewes2021, Raj2021}. Both routes need a strongly non-linear warp.
At $\alpha = H/r$, viscous damping keeps our discs out of that regime: at
$t = 1100\,\Porb$ the warp amplitude is at or below the non-linear threshold
$\psi \approx 1$ \citep{Held2026} in every run
(Table~\ref{tab:disc_response}), and the resolved velocity fluctuations
add no significant transport of their own ($\alpha_{\mathrm{R}} \approx
10^{-3}$; Section~\ref{sec:warp_no_break}). Differential precession, which
would wind the warp up, also stops early: the inner-disc node stalls after
shifting by at most $\sim 100^\circ$ in every run except the transiently decoupling one (Section~\ref{sec:branches}). \citet{Bourne2024} do find breaks at $q = 1$, $i = 45^\circ$, in self-gravitating, gravito-turbulent discs whose transport is not set by a laminar $\alpha$.
The closest approach to breaking in our grid shows this suppression at work. The
$e = 0$, $i = 60^\circ$ disc crossed into formal partial decoupling near
$t \approx 464\,\Porb$, its warp reaching nearly three times the local
non-linearity threshold $\psi \approx 1$ \citep{Held2026}, opening a real gap
in the surface density at a contrast of $3.2$, and was then pulled back:
its misalignment decayed by more than half, and at the radius where the gap
had formed no density minimum survives at all
(Section~\ref{sec:warp_no_break}; Appendix~\ref{apndx:transients}). Viscous suppression at $\alpha = H/r$ therefore acts as a restoring mechanism, and we observe it pulling back a disc that had crossed the threshold. A comparable stabilisation has been reported in
the Bardeen--Petterson problem, where hydrodynamic structures can hold a
disc together beyond the point at which the warp equations cease to admit a
steady solution \citep{Nealon2021}. The mechanisms are distinct: there the
torque is frame dragging and the stabilising agent is spiral structure,
while here it is the outward migration and broadening of the alignment
front that Appendix~\ref{apndx:transients} follows in time. Our
azimuthally averaged diagnostics carry no information on non-axisymmetric
structure, so we cannot test whether spirals contribute here as well. In both cases, crossing a formal breaking criterion does not commit a disc to breaking. For discs with $\alpha \lesssim 0.01$ at this geometry,
breaking remains the empirical expectation \citep{Rabago2024}; the test of
Appendix~\ref{apndx:detector} gives a reproducible check of that
prediction.

\subsection{Implications for massive black-hole binaries}
\label{sec:disc_implications}

Gas-driven eccentricity pumping is a leading route by which massive
black-hole binaries could reach the gravitational-wave bands with
measurable eccentricity. In purely stellar-dynamical evolution the
eccentricity is set largely stochastically when the binary forms and need not
grow \citep{Gualandris2022, Rawlings2023}, whereas in 2D, coplanar discs drive binaries toward
$e_{\mathrm{eq}} \approx 0.4$--$0.5$ \citep{Zrake2021, DOrazio2021, Siwek2023b}, with consequences for pulsar-timing-array residuals
\citep{Agazie2023} and for the eccentricity at LISA-band entry
\citep{AmaroSeoane2017_LISA}. Section~\ref{sec:orbital_maps} shows that
attractor to be inclination-dependent: it survives only at $i \lesssim
15^\circ$, and beyond $i \approx 30^\circ$ gives way to damping that
strengthens toward the polar runs. Massive black-hole binaries fed by
merger-driven \citep{Wang2025}, chaotically oriented gas inflows
\citep{King2006,Gaspari2013} sample the whole $(e, i)$ plane, so coplanar
eccentricity forecasts become upper bounds: only near-coplanar systems are
pumped, and gas hardening should deliver misaligned binaries to the
gravitational-wave bands more circular than 2D models predict. This holds for the prograde inclinations we sample: discs closer to retrograde that carry a few per cent of the binary mass can instead raise the eccentricity through von Zeipel--Kozai--Lidov-like oscillations and may shorten the merger time \citep{Martin2024}. Mass ratio
carries a comparable imprint of the accretion history: in coplanar 2D discs
the gas-driven approach to equal mass is slowest for highly eccentric
binaries already close to it, which need not reach $q = 1$ within a quasar
episode \citep{DeLaurentiis2026}. Whether misalignment speeds or slows that
evolution lies outside the reach of our equal-mass grid.

How the torque divides between the two channels adds a second, purely 3D
correction. Because a disc that aligns toward the binary plane delivers
almost all of its secular torque into reorienting the orbital plane
(Section~\ref{sec:phasemap}), a live misaligned binary would reorient its plane well before $\ab$ changed appreciably, while $\eb$ evolves on comparable timescales (Table~\ref{tab:timescales}), tracing a trajectory through the $(e, i)$ plane toward
one of the alignment attractors and integrating the response coefficients
measured here as it goes. Only where the disc is driven polar does the out-of-plane fraction fall, handing the torque to the in-plane channel
with the faster orbital shrinking and stronger circularisation that
Sections~\ref{sec:orbital_maps} and~\ref{sec:orbital_transient} document.

Table~\ref{tab:mbhb} translates the dimensionless timescales measured in
this paper into observing-frame periods for two anchor systems. At
milliparsec separations --- the regime where gas hardening operates,
where candidate periodic quasars are found \citep{DOrazioCharisi2023},
and where compact circumbinary discs are expected to grow observable
eccentric-disc asymmetries of their own \citep{Betancourt2026}
--- a $10^7\,\Msun$ binary has $\Porb \approx 0.3$ yr: the eccentric-forcing peak sits at months, the two-orbit alternation at $\sim 0.7$ yr and the lump at $\sim 1.5$ yr, so a decade of monitoring covers $\sim 30$, $15$ and $7$ cycles, the last marginal against red-noise variability \citep{DOrazioCharisi2023}; a polar-alignment transient lasts $\sim 30$ yr, longer than current survey baselines. At parsec separations every feature is secular: the
transient lasts of order a megayear, and the prediction becomes
demographic --- a calculable fraction of misaligned circumbinary systems
should be observed in the enhanced- or suppressed-accretion state, with reorienting,
possibly polar inner discs --- testable through imaging of cavity-scale
morphology and kinematics rather than through monitoring.

\begin{table}
\centering
\caption{Observing-frame timescales of the measured variability and
alignment features for two anchor systems, a $10^7\,\Msun$ binary at
$\ab = 0.5$ mpc ($\Porb \approx 0.33$ yr) and a $10^8\,\Msun$ binary at
$\ab = 1$ pc ($\Porb \approx 9.4 \times 10^3$ yr). Feature periods scale with $\Porb$ alone; the plane-reorientation time scales with the accretion time $M_{\rm b}/\langle\Mdot\rangle$ (Table~\ref{tab:timescales}), i.e. inversely with $\Mdot$ at fixed viscosity, and is quoted for the simulated disc.}
\label{tab:mbhb}
\small
\setlength{\tabcolsep}{3pt}
\begin{tabular}{lccc}
\toprule
feature & $P/\Porb$ & $10^7\,\Msun$, $0.5$ mpc & $10^8\,\Msun$, $1$ pc \\
\midrule
eccentric forcing & $1$ & $0.33$ yr & $9.4 \times 10^3$ yr \\
lump & $4.5$ & $1.5$ yr & $4.2 \times 10^4$ yr \\
polar transient & $\sim 100$ & $33$ yr & $\sim 1$ Myr \\
inner-disc precession & $\approx 435$ & $\sim 150$ yr & $\sim 4$ Myr \\
plane reorientation & $(1.5$--$11) \times 10^3$ & $0.5$--$3.5$ kyr & $14$--$100$ Myr \\
\bottomrule
\end{tabular}
\end{table}

A caveat accompanies the milliparsec column: at these separations the
binary is no longer Newtonian on secular timescales.
Gravitational-wave decay is the milder effect --- at the anchor's
parameters it proceeds on $\sim 10^8$ yr and leaves every tabulated
feature untouched. The consequential one is the 1PN apsidal precession of the binary, which
rotates the eccentricity vector $\hat{\vect{e}}_{\mathrm{b}}$ within the
orbital plane. Writing $P_{\mathrm{GR}}$ for the apsidal precession period
--- the time for $\hat{\vect{e}}_{\mathrm{b}}$ to complete one full
revolution --- $P_{\mathrm{GR}}/\Porb = c^2 \ab (1 - e_{\mathrm{b}}^2)/(3\,G
M_{\mathrm{b}}) \approx 2 \times 10^2$ at the milliparsec anchor for
$e_{\mathrm{b}} = 0.6$ (faster still for more eccentric binaries),
while at the parsec-scale anchor
$P_{\mathrm{GR}}/\Porb \approx 4 \times 10^4$ and the orbit is
effectively Keplerian. Two-dimensional simulations of coplanar discs
around precessing eccentric binaries show that this precession acts on the disc as well as on the orbit: for
$e_{\mathrm{b}} \gtrsim 0.25$ it imprints a quasi-sinusoidal
accretion-rate and light-curve modulation at the precession period,
second in periodogram power only to the orbital peak
\citep{DeLaurentiis2025}. At the milliparsec anchor that modulation lands in the same long-period range as the disc's own long-period signals of Section~\ref{sec:disc_observability}. They are distinguishable: the relativistic period follows from $M_{\mathrm{b}}$, $\ab$ and $e_{\mathrm{b}}$ alone, so it is predictable and can be cross-checked against a gravitational-wave measurement \citep{DeLaurentiis2025}, whereas the disc signals are set by the disc, and the one from a decoupling inner disc comes with a short-period signature the relativistic modulation lacks: lump suppression while the disc is strongly warped, and the lump's return as the interior aligns (Section~\ref{sec:disc_observability}).

For the alignment structure itself, the rotation of
$\hat{\vect{e}}_{\mathrm{b}}$ matters more than the variability: at
the milliparsec anchor $P_{\mathrm{GR}}$ is comparable to the
nodal-precession periods of the inner-disc gas that drives polar alignment, and where the eccentricity vector rotates faster than the disc can librate about
it, the polar state is effectively averaged away and the disc reverts to
circulation about $\lhat$.
The misaligned, three-dimensional response to a precessing binary is
unexplored --- \citet{DeLaurentiis2025} simulate coplanar discs and
flag even the coplanar back-reaction on the orbital evolution as open
--- so the alignment structure mapped here applies to relativistic
binaries only where the disc precession outpaces
$P_{\mathrm{GR}}$. A quantitative map of where polar alignment survives
relativistic apsidal precession in the $(e, i, \ab)$ space is a
natural extension of the fixed-orbit framework used here.

The response coefficients measured here are intended as input to a live-
binary calculation. Coupling $(\adot, \edot)$, the out-of-plane torque
fraction and the specific angular momentum delivered by accretion along
trajectories in the $(e, i)$ plane is the natural next step, with the
milliparsec systems of Table~\ref{tab:mbhb} as the observational target. Self-gravity \citep{Bourne2024}, magnetised
turbulence and radiative thermodynamics remain the principal physics
extensions.

\subsection{Polar discs around stellar binaries}
\label{sec:stellar_binaries}

Misaligned circumbinary discs are common around young stellar binaries, and the observed systems fall on the side of the critical inclination (Eq.~\eqref{eq:icrit}) that their state implies, as \citet{Murray2025} showed for well-characterised systems from \citet{Czekala2019}. Below it lie discs still tilted toward, and aligning with, the binary plane: KH~15D (a $\sim 15^\circ$ tilt at $\eb \approx 0.74$, where $i_{\rm crit} \approx 22^\circ$; \citealt{Smallwood2019}), GG~Tau~A ($30^\circ$ at $\eb = 0.45$, below $i_{\rm crit} \approx 42^\circ$; \citealt{Aly2018,Toci2024}) and IRS~43 ($> 60^\circ$ around a binary taken as circular, which has no polar state; \citealt{Brinch2016}). Above it lie polar discs: the protoplanetary disc of HD~98800~BaBb ($\eb \simeq 0.8$; \citealt{Kennedy2019,Franchini2019}), the debris ring of 99~Her \citep{Kennedy2012} and the dust disc of the post-AGB binary AC~Her \citep{Martin2023}. Protoplanetary discs are comparably thick (e.g. a gas scale height of about 7~au at 100~au in IRAS~04302+2247; \citealt{Villenave2023}) but far less turbulent, often $\alpha \lesssim 10^{-3}$ \citep{Flaherty2020}, and simulations of HD~98800~BaBb need $\alpha < 10^{-5}$ to match its low accretion rate \citep{Smallwood2022}. Such discs lie in the bending-wave regime, where breaking is the empirical expectation at our disc geometry (Section~\ref{sec:disc_regime}), as the torn disc of GW~Ori illustrates \citep{Bi2020,Young2025}. Where they stay coherent, $t_{\rm align} \propto \alpha^{-1}(H/r)^2$ (Section~\ref{sec:theory_warp_viscous}) stretches our polar half-times about fifty-fold at $\alpha = 10^{-3}$ and $H/r = 0.07$, still within a disc lifetime \citep{Lubow2018}.

The orbit evolves on the accretion time (Section~\ref{sec:orbital_maps}), and coplanar models need the binary to accrete at least 5--10 per cent of its mass for a significant change \citep{Valli2024}. Discs formed from the binary's own gas, after a common envelope \citep{Tuna2023} or in mass transfer \citep{Spruit2001}, should start nearly coplanar; misalignment needs gas from elsewhere, such as the natal cloud or an outer companion in a non-coplanar triple \citep{deVries2014,Chon2026}. Three implications follow for the polar systems above. First, the two outcomes proceed at different speeds: discs above $i_{\rm crit}$ reach the polar state about ten times faster than discs below it reach the binary plane (Table~\ref{tab:timescales}). A polar disc around a binary as eccentric as HD~98800~BaBb ($i_{\rm crit} \approx 19^\circ$) is therefore likely in its final state, whereas KH~15D, GG~Tau~A and IRS~43 are still in the slower alignment toward the binary plane. Second, the torque of a polar disc barely tilts the binary but damps its eccentricity and shrinks its orbit, faster per unit accreted mass than a coplanar disc; HD~98800~BaBb's surviving $\eb \simeq 0.8$ therefore suggests that its disc has supplied little mass, consistent with its low accretion rate. Third, accretion from a polar disc is suppressed, to $0.34$--$0.77$ of the coplanar rate (Fig.~\ref{fig:mdot_normalised}), but stays modulated at the orbital period (Fig.~\ref{fig:periodograms}), so accretion tracers such as H$\alpha$ should still vary with orbital phase. Accreting binaries with aligned discs lie on the coplanar equilibrium eccentricity \citep{Murray2025}, as our coplanar runs imply; in our runs it disappears once the disc is misaligned by $30^\circ$ or more (Section~\ref{sec:orbital_maps}).

\section{Conclusions}
\label{sec:conclusions}

We have presented a suite of fifteen 3D \arepo{} circumbinary disc
simulations spanning a binary eccentricity--inclination grid
($e = 0.0,\,0.3,\,0.6$; $i = 0^\circ$--$60^\circ$), with a locally isothermal $\alpha$-disc ($\alpha = H/r = 0.1$), a super-
Lagrangian refinement scheme, and a common analysis window ending at $t =
1100\,\Porb$ (Section~\ref{sec:methods_binary}). The suite follows the disc's
alignment evolution and the binary's orbital response within a single
measurement framework. Our main conclusions are as follows.

\begin{itemize}

\item \textit{Misaligned discs align in one of two ways, and binary
eccentricity selects between them.} Discs inclined by less than the
critical angle $i_{\rm crit}(e)$ (Eq.~\eqref{eq:icrit}) tilt gradually
back toward the binary orbital plane. Discs inclined by more rotate
their inner regions up to a near-perpendicular, polar orientation
($\delta \approx 88^\circ$). The measured outcomes agree with the
analytic prediction across the simulation grid; the two discs launched within
a few degrees of the transition, one on either side of it, barely change
their tilt, as expected where precession slows near the separatrix
(Figs~\ref{fig:delta_grid}, \ref{fig:mosaic} and \ref{fig:separatrix}).
Because $i_{\rm crit}$ decreases with binary eccentricity, the two
$i = 45^\circ$ discs end up on opposite sides of it: at fixed
inclination, eccentricity is what drives a disc toward coplanar
or polar alignment (Fig.~\ref{fig:phase_map}).

\item \textit{The disc's alignment dictates the binary's orbital response.}
Discs that align toward the binary plane channel nearly all of their secular torque into the out-of-plane component, driving precession of the orbital plane --- leaving only a small residual to affect $a$ and $e$ (Table~\ref{tab:disc_response}). In contrast, only when the inner disc fully reorients into the polar configuration does the in-plane torque --- responsible for driving $\dot{a}$ and $\dot{e}$ --- become dominant. At that point, the reorienting torque vanishes, and the same gas instead drives strong orbital shrinking and circularisation of
the binary orbit (Fig.~\ref{fig:orbital_maps}). The alignment pathway therefore sets both the disc's structure and how the disc couples to, and shapes, the binary's long-term evolution.

\item \textit{Discs settle before their binaries respond.} Polar alignment takes $124$--$312\,\Porb$, an order of magnitude faster than alignment toward the binary plane, and both lie within a factor of about three of the analytic alignment time (Table~\ref{tab:timescales}). The binary responds on the longer accretion time; around polar discs its inclination stays nearly fixed while its eccentricity changes 14--20 times faster and its semi-major axis about five times faster.

\item \textit{The turn to polar can lead to a transient and potentially observable event.} At $e=0.6$, $i=45^\circ$, the transition unfolds as a single coincident event: the accreted angular momentum reverses sign, the cavity shrinks to half its radius, the accretion rate rises to $\approx 2.7$ times its mean, and once the inner disc passes $\delta \approx 80^\circ$, the orbit shrinks 10--20 times faster. The more steeply inclined case at $i=60^\circ$ runs the same sequence faster and suppresses accretion to $0.3$ of its mean instead of enhancing it: the distance above $i_{\rm crit}$ sets both the speed and the character of the event (Appendix~\ref{apndx:transients}). This transition thus represents a dynamically rich, disc-driven phase in the binary's evolution --- where alignment, accretion, and orbital evolution are tightly coupled and potentially detectable.

\item \textit{In 3D, misalignment tends to circularise the binary.} For
circular binaries, residual eccentricity is damped at every inclination we
simulate, and most strongly at high inclination. The eccentricity pumping that
drives coplanar binaries toward $e \approx 0.4$ in 2D studies persists only up
to $i \approx 15^\circ$ and has reversed into damping by $i = 30^\circ$
(Fig.~\ref{fig:orbital_maps}). If this carries over to live binaries, gas-driven
binaries misaligned by up to $60^\circ$ may reach the gravitational-wave bands
less eccentric than coplanar models predict.

\item \textit{Accretion variability as a disc diagnostic.} The modulation at the orbital period, characteristic of eccentric binaries, persists through misalignment and even polar alignment. In contrast, the $\approx 4.5\,\Porb$ signal of circular binaries survives moderate misalignment but vanishes while a disc is strongly warped, reappearing as its interior realigns. Meanwhile, the disc that partially decouples modulates its accretion rate by about a quarter at the $\approx 435\,\Porb$ precession period of its inner disc, which then precesses independently of the outer disc. Thus, a periodogram of the accretion rate alone can constrain both the binary's eccentricity and the dynamical state of its disc.

\item \textit{Implications for misaligned stellar binaries.} Observed misaligned discs lie on the side of $i_{\rm crit}$ that their state implies: KH~15D, GG~Tau~A and IRS~43 below it, still aligning toward the binary plane, and HD~98800~BaBb and 99~Her above it, polar. Polar discs are likely in their final state; they barely tilt their binaries and feed them at a reduced rate that is still modulated at the orbital period (Section~\ref{sec:stellar_binaries}).

\end{itemize}

\medskip
\noindent\textit{Caveats and future work.} The binary orbit is fixed:
every rate reported here is a response coefficient measured at a
prescribed $(e, i)$, to be integrated along live trajectories rather
than read as an evolution track. Because the discs warp --- and in one run
transiently reach partial decoupling --- while the binary stays fixed,
our results describe the regime that precedes any global alignment
equilibrium, and a live-binary extension is needed to settle the
long-term fate of a partially decoupled configuration.
The disc is locally isothermal, non-self-gravitating, and viscous with
an imposed $\alpha$; live binaries, self-gravity, magnetised turbulence
and radiative thermodynamics are the natural physics extensions. Every disc
also starts with its node on the binary's apsidal meridian, where the polar
region is widest (Fig.~\ref{fig:circlib}b); away from it the critical
inclination rises, so a disc launched with a different node needs a larger
tilt to be driven polar. When the binary separation is only
$\sim 10^3\,G M_{\mathrm b}/c^2$ --- milliparsecs for a $10^7\,\Msun$
binary --- relativistic apsidal precession of the binary competes with the
disc dynamics, and mapping where polar alignment
survives it is a natural extension of this framework
(Section~\ref{sec:disc_implications}). Finally, whether a disc that recovers in this way can be driven to break by
a lower viscosity, a steeper density profile or a smaller cavity is the
natural follow-up our framework is set up to answer.

\section*{Acknowledgements}

LS would like to thank Taeho Ryu for numerous insightful discussions and
for his invaluable tutorials on computational hydrodynamics and the
\arepo{} code, which were foundational to this project and instrumental in
launching the author's journey in the field. LS is also grateful to Volker Springel for his support and trust throughout this
project, including access to the \arepo{} development repository, and for
reading the manuscript and providing helpful comments. Special thanks are due to
Oliver Zier for his technical support during the author's visit to the
Harvard-Smithsonian Center for Astrophysics (CfA), where he provided expert
guidance on code optimisation and simulation performance. Oliver also
offered a welcoming and intellectually stimulating environment, enabling
the presentation and discussion of the project's early scope and direction.
LS is further grateful to R\"{u}diger Pakmor for his invaluable advice on
the performance and convergence of the simulations.
Gratitude is also extended to Christopher Tiede for his fruitful and
motivating discussions on misaligned accretion discs around binary systems,
which took place during the author's visit to the Niels Bohr International
Academy (NBIA), and to Martin Pessah for hosting that visit. LS is also thankful to Hitesh Kishore Das for his
assistance with early-stage analysis and testing of key aspects of the code.
JC acknowledges financial support from ANID -- FONDECYT Regular 1251444, and Millennium Science Initiative Program NCN$2023\_002$.

\section*{Software}

\arepo{} \citep{Springel2010}, NumPy \citep{Harris2020}, SciPy \citep{Virtanen2020},
Matplotlib \citep{Hunter2007}, h5py \citep{Collette2013}.

\section*{Data Availability}

The data underlying this article will be shared on reasonable request to the
corresponding author.

\bibliographystyle{mnras}
\bibliography{main}

\appendix

\section{Refinement scheme}
\label{apndx:refinement}

\begin{figure}
\centering
\makebox[\columnwidth]{\includegraphics[width=0.9\columnwidth]{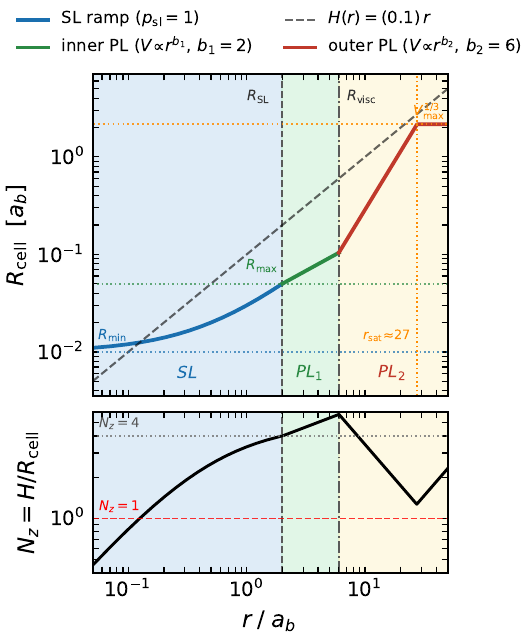}}
\caption{Target resolution of the super-Lagrangian refinement scheme versus spherical radius $r$ from the binary centre of mass. \emph{Top:} the effective Voronoi cell radius $r_{\mathrm{cell}}(r)$: the super-Lagrangian ramp inside $R_{\mathrm{SL}} = 2\,\ab$ (blue; $R_{\min} = 0.01$ to $R_{\max} = 0.05\,\ab$), the inner power law $V \propto r^{b_1}$ with $b_1 = 2$ (green) to $\Rvisc = 6\,\ab$, and the outer power law $V \propto r^{b_2}$ with $b_2 = 6$ (red), capped at $V_{\max}^{1/3} = 2.15\,\ab$ beyond $r_{\mathrm{sat}} \approx 27\,\ab$. The dashed line is the scale height $H = 0.1\,r$. \emph{Bottom:} cells per scale height, $N_z = H/r_{\mathrm{cell}}$: $N_z \gtrsim 4$ across the cavity wall and inner disc, peaking at $5.8$ at $\Rvisc$ (grey dotted line: $N_z = 4$), and never below $1.3$ in the disc body (red dashed line: $N_z = 1$). The blue branch averages the near-sink cell sizes azimuthally.}
\label{fig:refinement}
\end{figure}

The central numerical challenge in 3D CBD simulations is to resolve the cavity and
the cavity-wall gas --- where the binary--disc torque coupling is strongest ---
while keeping the cell count tractable across the far larger disc body. Standard
quasi-Lagrangian refinement holds the cell \emph{mass} fixed and therefore severely
under-resolves the low-density cavity. We instead impose a geometric
\emph{super-Lagrangian} (SL) target cell volume set by proximity to the sinks and
to the binary barycentre, following the scheme introduced for accreting black holes
by \citet{Curtis2015, Curtis2016} and adapted to live-binary CBDs by
\citet{Bourne2024}. Since the criterion acts on cell volume, we quote an effective
cell radius $r_{\mathrm{cell}} \equiv V_{\mathrm{cell}}^{1/3}$.

Within an SL sphere of radius $R_{\mathrm{SL}} = 2\,\ab$ centred on each sink, the
target cell radius ramps with distance $d$ to the nearest sink,
\begin{equation}
\label{eq:refine_sl}
r_{\mathrm{cell}}^{\mathrm{tar}}(d) = R_{\min}
   + (R_{\max} - R_{\min})\left(\frac{d}{R_{\mathrm{SL}}}\right)^{p},
\qquad d < R_{\mathrm{SL}},
\end{equation}
from $R_{\min} = 0.01\,\ab$ at the sink to $R_{\max} = 0.05\,\ab$ at $R_{\mathrm{SL}}$,
with ramp exponent $p = 1$ (the linear \citeauthor{Curtis2015} form). Outside the
sphere the target \emph{volume} follows a broken power law in the barycentric
radius $r$, anchored continuously to $V_{\mathrm{match}} = R_{\max}^3$ at
$r = R_{\mathrm{SL}}$:
\begin{equation}
\label{eq:refine_outer}
V_{\mathrm{cell}}^{\mathrm{tar}}(r) =
\begin{cases}
V_{\mathrm{match}}\left(r/R_{\mathrm{SL}}\right)^{b_1}, & R_{\mathrm{SL}} \le r \le \Rvisc, \\[4pt]
V_{\mathrm{visc}}\left(r/\Rvisc\right)^{b_2},          & r > \Rvisc,
\end{cases}
\end{equation}
with inner slope $b_1 = 2$, break radius $\Rvisc = 6\,\ab$, outer slope $b_2 = 6$,
and $V_{\mathrm{visc}} = V_{\mathrm{match}}(\Rvisc/R_{\mathrm{SL}})^{b_1}$ enforcing
continuity at the break. The target is capped at a background volume $V_{\max} = 10$
(i.e.\ $r_{\mathrm{cell}} \le 2.15\,\ab$), reached at $r_{\mathrm{sat}} \approx 27\,\ab$.
A cell is split when its volume exceeds $2\,V_{\mathrm{cell}}^{\mathrm{tar}}$ and
merged when it falls below $\tfrac{1}{2}V_{\mathrm{cell}}^{\mathrm{tar}}$. The
inner law is sink-centred and the outer law barycentre-centred, so the two coincide
for a single central mass and differ only within $\sim\!\ab$ of the binary; the SL
overlay switches on at $t = 10\,\Porb$ to let the initial mesh relax, a cavity clamp
holding inner cells at $V_{\mathrm{match}}$ beforehand. Unlike zone-based nested
grids, the prescription is seam-free: $r_{\mathrm{cell}}$ varies smoothly across the
whole domain.

Figure~\ref{fig:refinement} shows the resulting resolution. The cavity wall and inner
disc are covered at $N_z \equiv H/r_{\mathrm{cell}} \gtrsim 4$ cells per scale
height, peaking at $N_z \approx 5.8$ at $\Rvisc$, and the disc body stays resolved
at $N_z \gtrsim 1.3$ out to the disc edge; the minimum cell radius near the sinks is
$0.01\,\ab$, placing ${\sim}5$ cells across the accretion radius
$r_{\mathrm{acc}} = 0.05\,\ab$. Because the super-Lagrangian branch is referenced to
$d$ (the distance to the nearest sink) while the figure abscissa is the barycentric
radius, that branch ($r < R_{\mathrm{SL}}$) is plotted under the approximation
$d \approx r$ and represents the azimuthally-averaged near-sink resolution; the
$R_{\min}$ floor physically follows each component around its orbit.

We collect here the remaining implementation details and the production
parameter values (Table~\ref{tab:refinement_params}). The criterion acts on cell volume, with target
$v_{\mathrm{target}} = (r_{\mathrm{cell}}^{\mathrm{tar}})^3$ --- a cubed effective
radius rather than $\tfrac{4\pi}{3}r^3$, so that the accept/reject test is a direct
volume comparison. Within the super-Lagrangian sphere the controlling distance is
$d$, the 3D distance to the \emph{nearest} sink (Eq.~\eqref{eq:refine_sl}); outside it
the target is a function of the barycentric radius $r$ (Eq.~\eqref{eq:refine_outer}),
anchored continuously at $V_{\mathrm{match}} = R_{\max}^3$. The two distance metrics
coincide for a single central mass and differ by up to ${\sim}\ab$ near the binary,
where each component orbits at $\approx 0.5\,\ab$ from the barycentre ($q = 1$);
accordingly the super-Lagrangian branch of Fig.~\ref{fig:refinement} is shown against
$r$ using $d \approx r$ and is an azimuthal average of the near-sink cell sizes.

A cell is refined (split) when its volume exceeds $2\,v_{\mathrm{target}}$ and
de-refined (merged) when it falls below $\tfrac{1}{2}\,v_{\mathrm{target}}$; cells
with volume below $10^{-2}\,V_{\mathrm{floor}}$ are merged unconditionally to suppress
pathological near-sink cells. The super-Lagrangian overlay is inactive for
$t < 10\,\Porb$, during which a cavity clamp holds inner cells at $V_{\mathrm{match}}$
while the initial mesh relaxes.

\begin{table}
\centering
\renewcommand{\arraystretch}{1.2}
\caption{Refinement-scheme parameters used in all production runs.}
\label{tab:refinement_params}
\begin{tabular}{lll}
\toprule
Parameter & Symbol & Value \\
\midrule
\texttt{SinkRefRadius}        & $R_{\mathrm{SL}}$              & $2.0\,\ab$ \\
\texttt{SinkRefMaxCellRadius} & $R_{\max} = r_{\mathrm{cell}}(R_{\mathrm{SL}})$ & $0.05\,\ab$ \\
\texttt{SinkRefMinCellRadius} & $R_{\min} = r_{\mathrm{cell}}(0)$    & $0.01\,\ab$ \\
\texttt{PowerLawSL}           & $p$                            & 1.0 \\
\texttt{R\_visc}              & $\Rvisc$                       & $6.0\,\ab$ \\
\texttt{PowerLawInner}        & $b_1$                          & 2.0 \\
\texttt{PowerLawOuter}        & $b_2$                          & 6.0 \\
\texttt{SLStartTime}          & ---                            & $10\,\Porb$ \\
\texttt{MaxVolume}            & $V_{\max}$                     & $10$ ($r_{\mathrm{cell}} \le 2.15\,\ab$) \\
\texttt{CavityFloorVolume}    & $V_{\mathrm{floor}}$           & $10^{-5}$ \\
\bottomrule
\end{tabular}
\end{table}

\section{Supplementary disc structure and accretion}
\label{apndx:gallery}

This appendix collects per-run material that supports Sections~\ref{sec:disc_response}
and \ref{sec:discussion} without carrying their arguments: the in-plane gas
structure of every run at a common epoch, and the accretion histories behind
the mean rates quoted in Section~\ref{sec:disc_observability}.

\subsection*{Face-on gas structure}

\begin{figure*}
\makebox[\textwidth]{\includegraphics[width=0.85\textwidth]{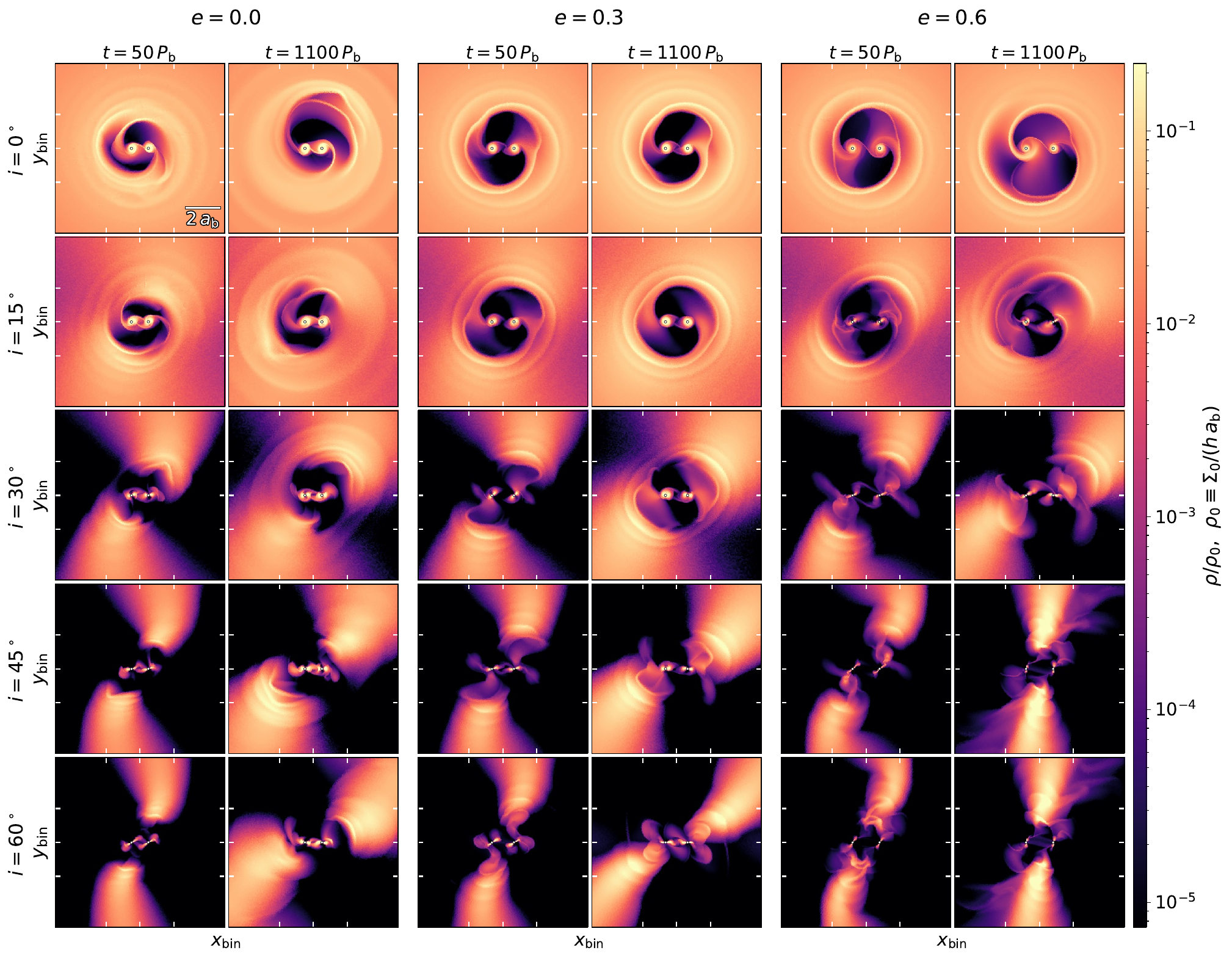}}
\caption{Gas density in the binary orbital plane for all fifteen runs at $t = 50$ and $1100\,\Porb$ (rows: inclination; column pairs: eccentricity): nearest-cell mid-plane cuts in the binary frame over a $10 \times 10\,\ab$ field of view, on a shared logarithmic colour scale in units of $\rho_0 = \Sigma_0/(h\,\ab)$; white dots mark the sinks and the bar is $2\,\ab$.  From $i = 30^\circ$ the disc intersects the binary plane along its line of nodes; the bright band is that intersection, so cavity size is not comparable across inclination in this projection (Fig.~\ref{fig:phase_map} measures it in 3D). At $i = 30^\circ$ ($e = 0$ and $0.3$) the crossing band of $t = 50\,\Porb$ has become a face-on disc by $1100\,\Porb$, as the innermost disc flattens into the binary plane.}
\label{fig:xy_mosaic}
\end{figure*}

Figure~\ref{fig:xy_mosaic} shows the gas density in the binary orbital plane
for all fifteen runs, at the first and last epochs of Fig.~\ref{fig:mosaic}
($t = 50$ and $1100\,\Porb$) so that the two figures may be read together.
The rendering is a \emph{mid-plane cut}: the volume
density is interpolated onto the plane by nearest-cell Voronoi sampling ---
the same routine used for Fig.~\ref{fig:mosaic}, applied to the orthogonal
plane --- over a $10 \times 10\,\ab$ field of view, on a logarithmic colour
scale in $\rho/\rho_0$ with $\rho_0 \equiv \Sigma_0/(h\,\ab)$ shared across
all thirty panels.

The plane is the binary's own, for every run. We cut along this plane because it is the one in which the binary orbits, in which the cavity is
carved, onto which the accretion streams land, and in which any
cavity-edge overdensity circulates. Because the operation is identical for
every run, the comparison across the grid is consistent, and the progressive
loss of material from the panels at high inclination is a physical statement
--- the disc has largely left the binary's plane --- rather than a projection
artefact. Two things follow for how the figure should be read. From
$i = 30^\circ$ the disc intersects the plane along its line of nodes, so each
panel shows a crossing band, narrowing as inclination grows, and the dark
region around it is a geometric absence rather than an evacuated cavity.
Consequently cavity \emph{size} may not be compared across inclination here;
that quantity is measured in three dimensions and is carried by the
$\langle r_{20}\rangle$ panel of Fig.~\ref{fig:phase_map}. What this figure
carries is morphology.

At $i = 0^\circ$ and $15^\circ$ the panels are textbook face-on circumbinary
discs: a cleared cavity, a two-armed spiral, and two streams feeding the
sinks, with the cavity growing larger and visibly more lopsided as
eccentricity increases. From $i = 30^\circ$ the crossing band dominates,
though the circumsingle discs and the stream footpoints remain resolved near
the sinks; at $e = 0.6$ and $i = 45^\circ$, $60^\circ$ the band stands nearly
vertical by $t = 1100\,\Porb$, which is polar alignment
(Section~\ref{sec:branches}) seen in the binary's own frame.

The exceptions are the $i = 30^\circ$ discs at $e = 0$ and $0.3$: their $t = 50\,\Porb$ panels are ordinary crossing bands, while by $t = 1100\,\Porb$ both show a face-on disc with a cavity and spiral arms, because their innermost rings have flattened into the binary's plane. The alignment front of Section~\ref{sec:branches} is thus directly visible in the density field, with the still-inclined exterior forming the band around the flattened interior.

\begin{figure*}
\includegraphics[width=\textwidth]{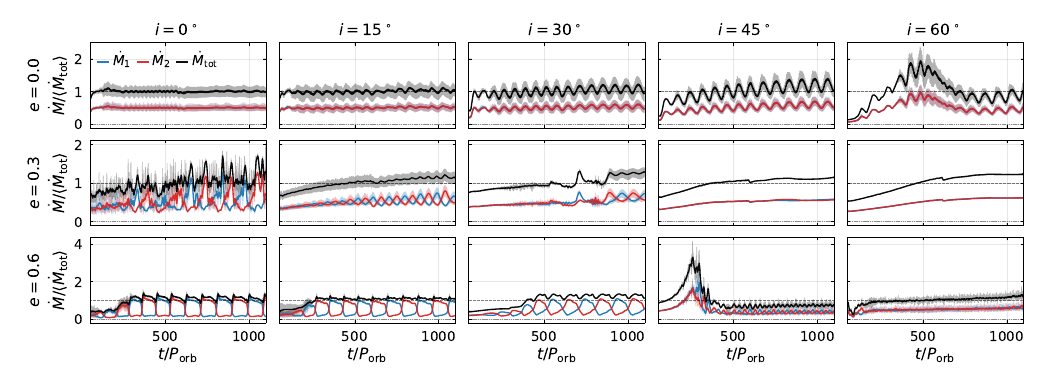}
\caption{Accretion rates over the common window $t = 50$--$1100\,\Porb$ for
all fifteen runs (rows: eccentricity; columns: inclination), normalised by
each run's window-mean total rate: $\Mdot_1$ (blue), $\Mdot_2$ (red) and
$\Mdot_{\rm tot}$ (black). Light curves are sampled every $\Porb$, bold
curves smoothed over $10\,\Porb$.}
\label{fig:mdot_evolution}
\end{figure*}

\subsection*{Accretion histories and mean rates}

Figure~\ref{fig:mdot_evolution} shows the accretion rate onto each sink and
onto the binary as a whole, over the common window $t = 50$--$1100\,\Porb$,
normalised by each run's own window mean. The individual-sink curves make the
preferential-accretion behaviour visible --- the $\Mdot_1 \leftrightarrow
\Mdot_2$ alternation in the eccentric coplanar runs on the $167$ and $174\,\Porb$ precession periods of their eccentric cavity edge, as \citet{Dunhill2015} found for the eccentric binary HD~104237 --- and the transient events discussed in Section~\ref{sec:orbital_transient}
appear directly as the $2.7\times$ rise at $e=0.6$, $i=45^\circ$ and the
suppression at $e=0.6$, $i=60^\circ$.

The mean rates over $t = 100$--$1100\,\Porb$ condensed from these histories are plotted in
Fig.~\ref{fig:mdot_normalised} and discussed in
Section~\ref{sec:disc_observability}.

\section{Time-resolved evolution of the transient episodes}
\label{apndx:transients}

While most discs in the simulation grid warp and align gradually, in two cases the disc passes through a transient episode: the partial decoupling of the $e=0$, $i=60^\circ$ disc (Section~\ref{sec:warp_no_break}) and the turn to polar of the $e=0.6$, $i=45^\circ$ disc, which drags the binary's orbital response with it (Section~\ref{sec:orbital_transient}). This appendix gives the time-resolved measurements behind the main-text summaries.

\subsection*{The transient decoupling at $e=0$, $i=60^\circ$}

\begin{figure}
\includegraphics[width=\columnwidth]{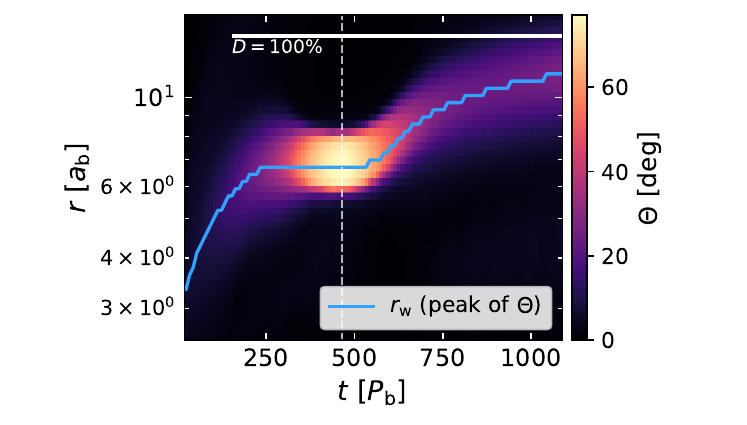}\\[3pt]
\makebox[\columnwidth]{\includegraphics[width=0.75\columnwidth]{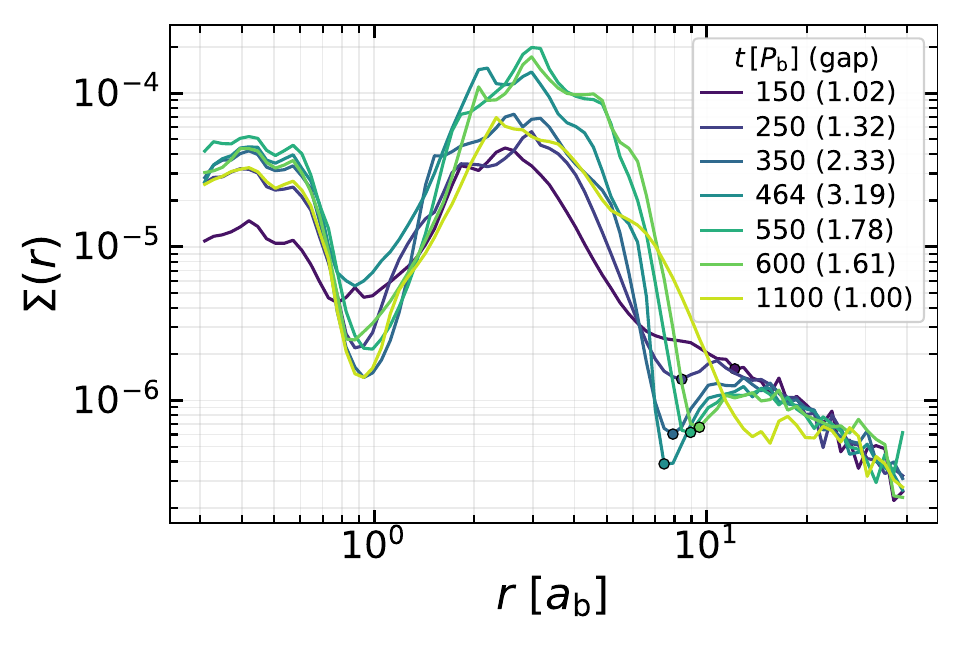}}
\caption{The transient partial decoupling of the $e=0$, $i=60^\circ$ disc, in sliding $25\,\Porb$ windows. \textit{Top:} the inner--outer misalignment $\Theta(r,t)$ (Eq.~\eqref{eq:theta}, reference baseline as in Fig.~\ref{fig:warp_vs_break}), with the warp radius; the white bar marks the interval over which the duty cycle, the fraction of each window in which $\Theta$ at the warp radius exceeds $15^\circ$ (Appendix~\ref{apndx:detector}), is full; the dashed line marks peak misalignment, $t \approx 464\,\Porb$. \textit{Bottom:} the surface-density profile at seven epochs, with the gap contrast (legend) measured at the radius where the gap opened ($r = 6.7\,\ab$); markers locate the density minima, absent at $t = 1100\,\Porb$, when the gap has refilled. }
\label{fig:theta_rt}
\end{figure}

This disc built the largest warp of the grid --- $\psi_{\rm eff}$ reached
$2.71$ at its peak, before relaxing to $0.82$ by $t = 1100\,\Porb$ --- and
is the only disc to reach partial decoupling at all. Its
interior and exterior are each internally coherent yet hold a sustained
mutual offset, co-located with a surface-density gap and a local disc-
eccentricity excess. Figure~\ref{fig:theta_rt} follows the feature in time.
It is detectable from $t \approx 134\,P_{\rm b}$ and reaches full duty (misaligned beyond $15^\circ$ throughout each window; Appendix~\ref{apndx:detector}) by $t \approx 154\,P_{\rm b}$; over the next $\sim$300 orbits it migrates outward
while \emph{narrowing}, its width falling from $0.91$ to $0.34$ in $\ln r$. It peaks at $t \approx 464\,P_{\rm b}$, where the two sides are
misaligned by $\Theta = 77^\circ$ across an interface
evacuated to a density contrast of $3.2$, with the feature at $r =
6.7\,a_{\rm b}$ and at its sharpest --- formally a partial decoupling, two
coherent discs separated by a gradient too steep to be a simple warp.

It does not complete. Amplitude, sharpness and evacuation turn over together
at that epoch and decline thereafter: by the end of the run the misalignment
has fallen by $64$ per cent to $\Theta = 28^\circ$, the feature has migrated
$70$ per cent further out to $r = 11.4\,a_{\rm b}$ and broadened again, and
the density gap has refilled completely where it opened, leaving only a shallow residual minimum near $r \approx 15\,\ab$. The duty cycle never falls below $99$ per cent: the disc stays continuously misaligned while the misalignment decays.
The mechanism is visible in the tilt profiles of
Fig.~\ref{fig:delta_grid}: the alignment front that produced the steep
gradient migrates outward and broadens as it travels, spreading the
accumulated misalignment over more scale lengths than it can sharpen, while
the interior it leaves behind passes through its most-aligned state
(reaching $\delta \approx 9^\circ$ near $t \approx 594\,P_{\rm b}$) and then
tilts back to $\approx 33^\circ$.

\subsection*{The polar-alignment transient at $e=0.6$, $i=45^\circ$}

\begin{figure}
\makebox[\columnwidth]{\includegraphics[width=0.7\columnwidth]{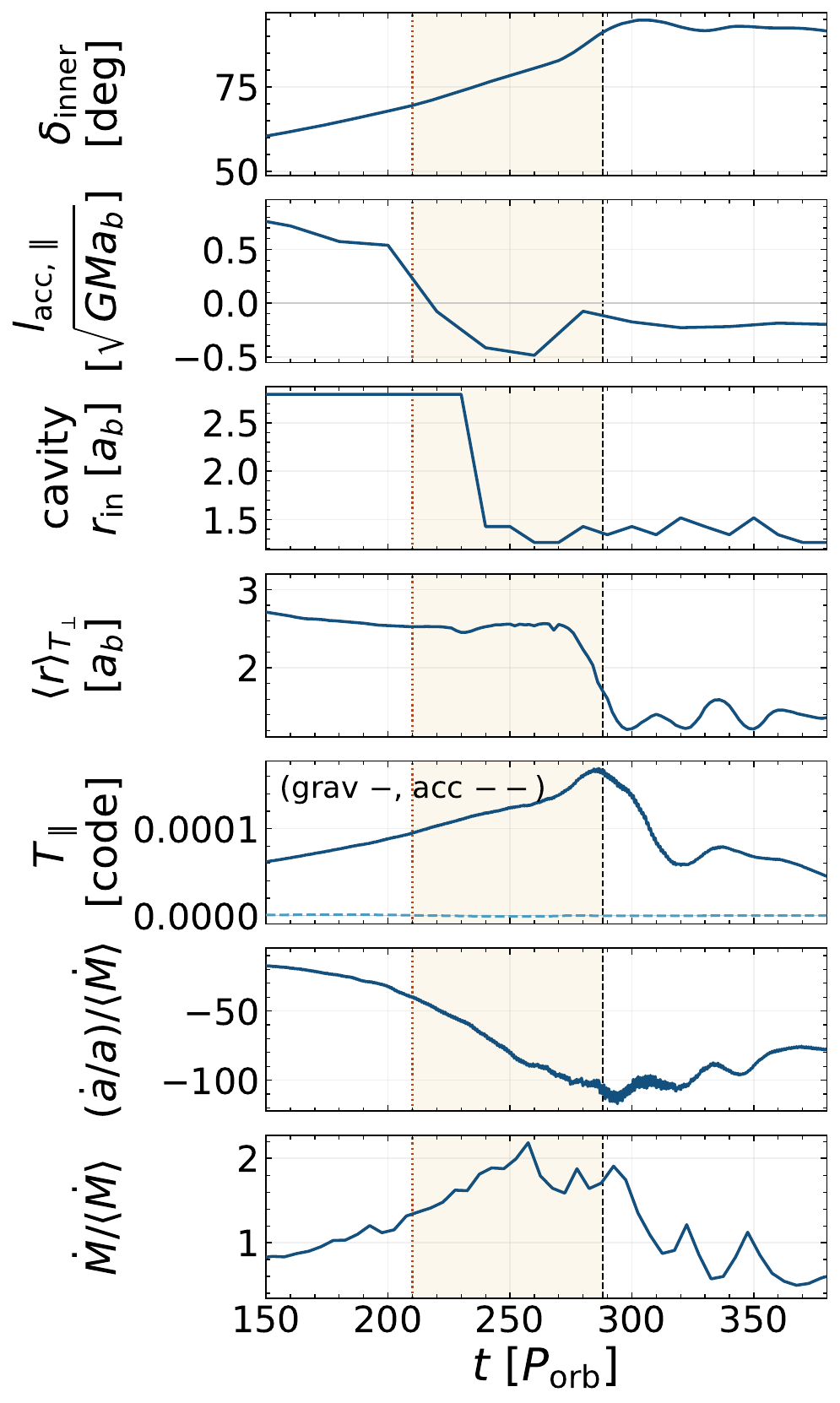}}
\caption{The polar-alignment transient at $e=0.6$, $i=45^\circ$, shown over $t = 150$--$380\,\Porb$. From top: inner-disc tilt $\delta$ ($r = 2$--$4\,\ab$); the change in the binary's relative specific angular momentum per unit accreted mass, $l_{\mathrm{acc},\parallel}$ (Section~\ref{sec:orbital_transient}); cavity $\Sigma$-peak radius; radial centroid $\langle r \rangle_{T_\perp}$ of the precession torque; in-plane torque $T_\parallel$ (gravitational solid, accretion dashed, $\sim 100\times$ smaller); $(\adot/\ab)/\langle\Mdot\rangle$; and $\Mdot/\langle\Mdot\rangle$. The amber dotted line marks the $l_{\mathrm{acc},\parallel}$ sign flip ($t \approx 210\,\Porb$), the black dashed line the drop of the out-of-plane fraction ($t \approx 288\,\Porb$), and the shaded band the transient between them. }
\label{fig:transient_stack}
\end{figure}

At $e = 0.6$, $i = 45^\circ$ the reorientation of Section~\ref{sec:branches} is a single dynamical event in which the accretion flow, the cavity, the out-of-plane torque fraction and the orbital-element response reorganise together. Figure~\ref{fig:transient_stack} stacks the relevant diagnostics on
one time axis. The causal ordering is well resolved:
\begin{enumerate}
\item the specific angular momentum delivered by accretion,
      $l_{\mathrm{acc},\parallel}$, after an early prograde rise to
      $\approx +0.9$, crosses zero at $t \approx 210\,\Porb$
      (inner-disc tilt $\delta \approx 70^\circ$) and settles retrograde at
      $\approx -0.25$ --- the earliest marker of the transition;
\item the cavity shrinks, its $\Sigma$-peak radius dropping from
      $2.8$ to $1.3\,\ab$ over $t \approx 230$--$250\,\Porb$;
\item the total accretion rate rises to $\approx 2.7$ times its secular
      mean at $t \approx 255\,\Porb$;
\item the precession-torque radial centroid $\langle r \rangle_{T_\perp}$
      moves in to the cavity edge and the out-of-plane fraction drops
      to its polar-state value at $t \approx 288\,\Porb$, coincident with
      the deepest excursion of $\adot/\ab$,
      $(\adot/\ab)/\langle\Mdot\rangle \approx -115$, once the inner-disc
      tilt crosses $\delta \approx 80$--$85^\circ$.
\end{enumerate}
The out-of-plane fraction responds late: $\delta$ ramps smoothly for $\sim 150$ orbits while $\langle r \rangle_{T_\perp}$ holds steady, and the fraction drops only near $\delta = 90^\circ$.

The in-plane torque panel of Fig.~\ref{fig:transient_stack} identifies the
driver. The gravitational component $T_{\parallel,\mathrm{grav}}$ rises
smoothly through the transient and peaks exactly as the out-of-plane fraction drops; the direct
accretion torque is $\sim 100$ times smaller throughout. The sign flip of the delivered angular momentum thus \emph{traces} the stream reorientation, while gravity does the dynamical work. The torque is positive while $\ab$ decreases because circularisation dominates the
angular-momentum budget at $e_{\mathrm{b}} = 0.6$ (Eqs.~\eqref{eq:adot}--\eqref{eq:edot}).

The second polar run ($e = 0.6$, $i = 60^\circ$) goes through the same sequence faster, retrograde from the start and with its accretion rate suppressed rather than enhanced (Section~\ref{sec:orbital_transient}). Only the two polar runs show such a transient: their smoothed secular rates span $\approx 105$ peak to peak, stable to the smoothing, against $4$--$8$ in every other run (not shown).

\section{Testing the discs for breaks}
\label{apndx:detector}

\begin{figure}
\makebox[\columnwidth]{\includegraphics[width=0.75\columnwidth]{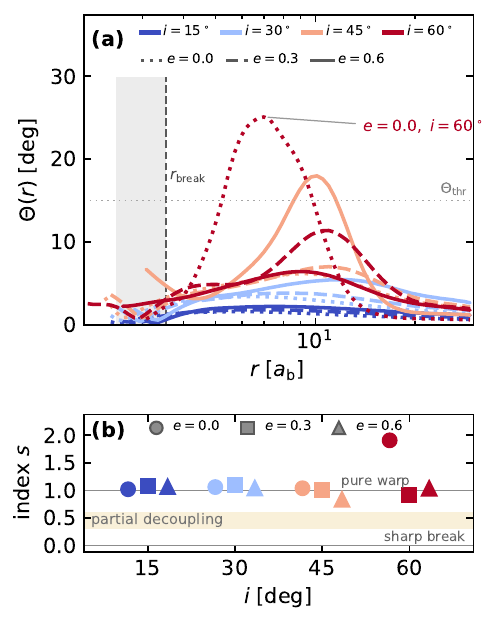}}
\caption{The break test applied to the twelve inclined runs.
(a) Mutual misalignment $\Theta(r)$ (Eq.~\eqref{eq:theta}, reference baseline $h=0.35$): colour darkens with inclination, linestyle
gives eccentricity. (b) The baseline scaling index $s$ (Eq.~\eqref{eq:scale_index}), median over $t = 11$--$1100\,\Porb$, for each run: $s \rightarrow 1$ identifies a
smooth warp ($\Theta \propto h$), $s \rightarrow 0$ a sharp break ($\Theta$
independent of $h$); the shaded band marks the
partial-decoupling range. The grey band in panel (a) is the tearing radius of Eq.~\eqref{eq:r_break_nixon} across the grid's inclinations ($2.50$--$3.54\,\ab$; dashed line at its $i = 45^\circ$ maximum), inside every measured warp radius.}
\label{fig:warp_vs_break}
\end{figure}

Visual inspection of $\delta(r)$ or $\Sigma(r)$ cannot reliably tell a broken disc (two planes separated by a discontinuity) from a strong but continuous warp; we separate them as follows.

At every candidate radius we measure the mutual misalignment between the
disc interior and exterior to $r$,
\begin{equation}
\Theta(r) \;=\; \arccos\!\bigl[\hat{\vect{l}}_{\rm in}(r)\cdot
\hat{\vect{l}}_{\rm out}(r)\bigr],
\label{eq:theta}
\end{equation}
where $\hat{\vect{l}}_{\rm in}$ and $\hat{\vect{l}}_{\rm out}$ are the
directions of the \emph{summed} shell angular momenta
$\sum_k \vect{L}_{\mathrm d}(r_k)$ over the logarithmic windows
$r/f \leq r_k < r$ and $r \leq r_k < r f$, with $f = e^{h}$; the reference
half-width is $h = 0.35$. Summing $\vect{L}$ before normalising weights each
window by $|\vect{L}|$, so near-empty cavity bins cannot dominate, and
$\Theta$ remains well defined for a polar disc whose node is degenerate. The
same weighting makes each window dominated by its outer edge, so the warp
radius $r_{\rm w}$ --- the radius that maximises $\Theta$ ---
underestimates the radius of the transition it locates.

A break is a discontinuity, so $\Theta$ does not grow with the baseline; a
smooth warp is a gradient, so $\Theta \propto h$. The scaling index
\begin{equation}
s \;=\; \frac{\Theta(h_1)/\Theta(h_0) \,-\, 1}{h_1/h_0 \,-\, 1},
\label{eq:scale_index}
\end{equation}
evaluated at $r_{\rm w}$, compares a narrow and a wide baseline and gives $s \rightarrow 1$ for a warp of uniform gradient and $s \rightarrow 0$ for a break; $s > 1$ means the gradient steepens away from $r_{\rm w}$, as in a broad, curved warp. We use
$h_0 = 0.15$, which spans $3$--$4$ shells to each side at the shell
resolution $\Delta\ln r \simeq 0.04$, and $h_1 = 0.35$, which still lies
inside the measured transitions ($\Delta\ln r \approx 0.4$--$0.5$), where a
warp is still growing but a discontinuity has saturated. The small-baseline
slope of the same scan gives the effective warp amplitude, our measured
proxy for $\psi$ (Eq.~\eqref{eq:psi}),
\begin{equation}
\psi_{\rm eff} \;=\; \lim_{h \to 0} \frac{\Theta(h)}{h}
\;\simeq\; \frac{\Theta(h_0)}{h_0}
\;=\; \left|\frac{{\rm d}\delta}{{\rm d}\ln r}\right|_{r_{\rm w}}
\label{eq:psieff}
\end{equation}
(in radians), and the duty cycle $D$ is the fraction of the analysis window
during which the instantaneous $\Theta(r_{\rm w}, t)$ exceeds
$\Theta_{\rm thr} = 15^\circ$. Two guards keep the test reliable: each side of a trial radius must hold a steady orientation across the window, and radii whose shells are too sparse to define a direction, at the cavity and the outer edge, are excluded.

Figure~\ref{fig:warp_vs_break} applies the test to the twelve inclined
runs. Every misalignment profile peaks in the disc body, and eleven of the
twelve runs sit at $s = 0.83$--$1.10$, essentially on the pure-warp scaling
$\Theta \propto h$, both polar discs among them. The twelfth, $e = 0$,
$i = 60^\circ$, has $s = 1.91$: its misalignment grows faster than in
proportion to the baseline, still a warp and not a discontinuity. These values are medians over $t = 11$--$1100\,\Porb$; at the peak of its partial decoupling ($t \approx 470\,\Porb$) the $e = 0$, $i = 60^\circ$ interface is too depleted for the test to resolve, and Fig.~\ref{fig:theta_rt} follows that episode in time. No run falls in the
partial-decoupling range, let alone near the break limit.

\bsp
\label{lastpage}
\end{document}